\documentclass[10pt,twocolumn,letterpaper]{article}

\usepackage[pagenumbers]{cvpr} 

\usepackage[dvipsnames]{xcolor}
\newcommand{\red}[1]{{\color{red}#1}}

\newcommand{\authorskip}{\hspace{5mm}}
\usepackage{listings}

\definecolor{tablegreen}{RGB}{11,225,83}
\definecolor{tablered}{RGB}{230,28,100}
\definecolor{tableblue}{RGB}{60,105,225}
\definecolor{tablegray}{gray}{0.45}

\newlength\savedwidth
\newcommand\whline{\noalign{\global\savedwidth\arrayrulewidth\global\arrayrulewidth 0.8pt}
\hline\noalign{\global\arrayrulewidth\savedwidth}}

\newcommand{\algname}{MAN}

\usepackage{soul}

\usepackage[ruled,linesnumbered]{algorithm2e}
\usepackage{amsmath,amsfonts,bm}

\def\eqref#1{equation~\ref{#1}}

\def\1{\bm{1}}

\DeclareMathAlphabet{\mathsfit}{\encodingdefault}{\sfdefault}{m}{sl}
\SetMathAlphabet{\mathsfit}{bold}{\encodingdefault}{\sfdefault}{bx}{n}

\usepackage{multirow}
\usepackage[export]{adjustbox}
\usepackage{svg}
\usepackage{makecell}

\definecolor{tablegreen}{RGB}{235,255,210}
\definecolor{tablered}{RGB}{255,255,220}
\definecolor{tableblue}{RGB}{240,240,255}

\definecolor{Highlight}{HTML}{39b54a}  %
\definecolor{red}{HTML}{ff3509} 
\definecolor{blue}{RGB}{15,45,245}
\definecolor{green}{RGB}{25,130,45}
\definecolor{lightgray}{gray}{0.9}%

\usepackage{appendix}
\usepackage{pifont}
\usepackage{xcolor}
\usepackage{colortbl}
\renewcommand{\red}[1]{\textbf{#1}}
\newcommand{\blue}[1]{\underline{#1}}

\usepackage{arydshln}

\usepackage{svg}

\definecolor{cvprblue}{rgb}{0.21,0.49,0.74}
\usepackage[pagebackref,breaklinks,colorlinks,citecolor=cvprblue]{hyperref}
\usepackage[accsupp]{axessibility} 

\def\paperID{17} 
\def\confName{NTIRE}
\def\confYear{2024}

\title{Multi-scale Attention Network for Single Image Super-Resolution}

\author{
	Yan Wang\footnotemark[2] \authorskip
        Yusen Li\footnotemark[2] \authorskip
	Gang Wang\authorskip
	Xiaoguang Liu 
        \\[2mm]
	Nankai-Baidu Joint Lab, Nankai University 
   \\[2mm]
{\tt Code:\ \url{https://github.com/icandle/MAN}} 
\vspace{-2mm}
}

\begin{document}
\maketitle

\renewcommand{\thefootnote}{\fnsymbol{footnote}}
\footnotetext[2]{Corresponding authors, email: \{wangy, yusenli\}@nbjl.nankai.edu.cn.}
\begin{abstract}
ConvNets can compete with transformers in high-level tasks by exploiting larger receptive fields. To unleash the potential of ConvNet in super-resolution, we propose a \underline{m}ulti-scale \underline{a}ttention \underline{n}etwork (MAN), by coupling classical multi-scale mechanism with emerging large kernel attention. In particular, we proposed multi-scale large kernel attention (MLKA) and gated spatial attention unit (GSAU). Through our MLKA, we modify large kernel attention with multi-scale and gate schemes to obtain the abundant attention map at various granularity levels, thereby aggregating global and local information and avoiding potential blocking artifacts. In GSAU, we integrate gate mechanism and spatial attention to remove the unnecessary linear layer and aggregate informative spatial context. To confirm the effectiveness of our designs, we evaluate MAN with multiple complexities by simply stacking different numbers of MLKA and GSAU. Experimental results illustrate that our MAN can perform on par with SwinIR and achieve varied trade-offs between state-of-the-art performance and computations.
\end{abstract}

\section{Introduction}
Image super-resolution (SR) is a widely concerned low-level computer vision task that focuses on rebuilding the missing high-frequency information from the low-quality input~\cite{sun2008image,tai2010super,he2011single,CAMixerSR}. However, it is ill-posed that one low-resolution (LR) image corresponds to countless potential high-resolution (HR) images, leading to difficulties in finding proper correlations between the LR and HR pixels. Due to the boom of deep neural networks, several CNN- and transformer-based SR models~\cite{SRCNN,FSRCNN,EDSR,RCAN,SwinIR} have been developed that use prior and intra-image information to improve the reconstruction quality. Generally, they approach the issue from two perspectives.

The first and simplest way is to enlarge the model capacity by training the network with larger datasets and better strategies. Specifically, based on ImageNet~\cite{ImageNet}, IPT~\cite{IPT} and HAT~\cite{HAT} conducted a sophisticated pre-training to excavate the capability of transformers in image processing. LSDIR~\cite{LSDIR} introduced a large-scale dataset to exploit model capacity fully. RCAN-it~\cite{RCAN_it} leveraged reasonable training strategies to help RCAN~\cite{RCAN} regain SOTA performance. Generally, these approaches are universal for neural models but increase burdensome training and data collection consumption.

\begin{figure}[!t]
\centering
\includegraphics[width=0.95\linewidth]{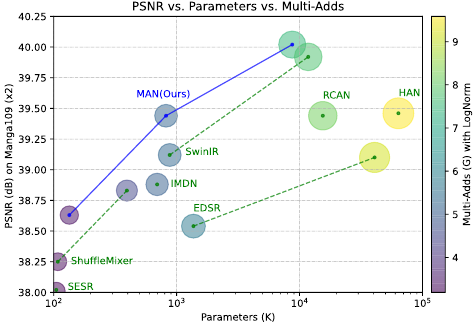} 
\caption{Trade-off between performance and model complexity on Manga109~\cite{Manga109} with $\times$2 SR scale. MANs can achieve higher PSNR with fewer parameters and computations. }
\label{fig:Network_Performance}
\end{figure}

Another effective way is to activate more intra-image information via better network design. One primary idea is to enlarge the perceptive fields, which means a deeper and wider network topology. Following this, \cite{VDSR,EDSR,RCAN} continuously expanded the network depth to hundreds of layers. Nevertheless, the improvement brought by this strategy is limited by over-training and high training costs. Recent models~\cite{RFANet,HDRN,ENLCA} have employed complex topology and attention mechanisms to capture more useful information, \eg, multi-scale~\cite{MSRN} design and non-local attention~\cite{NLSA}. 

\begin{table}[t]
\center
\begin{center}
\caption{SR Performance of ConvNets \emph{vs.} Transformer backbones.}
\label{tab:1}
\small
\fontsize{8.5pt}{9.5pt}\selectfont
\tabcolsep=2.5pt
\begin{tabular}{l|c|c|c|c|c|c}
\whline
{Backbone} &
{\#Params} & {\#FLOPs} &  {Set5} & {Set14}&  {B100} & {U100}   \\
\whline
ConvNeXt-S~\cite{ConvNeXt} & 833K & {47.3G} %
& {32.22}
& {28.62}
& {27.60}
& {26.07}
\\
VAN-S~\cite{VAN} & 818K & {46.0G} %
& {32.27}
& {28.70}
& {27.63}
& {26.18}
\\
SwinIR-light~\cite{SwinIR} & 896K & {49.6G} %
& {32.30}
& {28.73}
& {27.65}
& {26.30}
\\
\rowcolor{tableblue}
MAN-light & 840K & {47.1G} %
& {32.33}
& {28.76}
& {27.67}
& {26.31}
\\
\whline
\end{tabular}
\end{center}
\end{table}

Recently, the transformer-based models~\cite{ViT,IPT,SwinIR,EDT} have shown a remarkable representation ability of self-attention~(SA), which gradually superseded ConvNets as the state-of-the-art model in both high- and low-level tasks. To fight back, many pure ConvNet (ConvNeXt~\cite{ConvNeXt}, VAN~\cite{VAN}) sprouted and achieved comparable performance in high-level tasks. But do they perform well in low-level? In \cref{tab:1}, we tested them and noticed VAN performs better than ConvNeXt but still falls behind transformers. To modernize the VAN to compete with transformers in the SR field, we reassess the design of VAN. Generally, VAN~\cite{VAN} explored kernel decomposition and proposed large kernel attention (LKA), where a large kernel can be replaced by stacked depth-wise, depth-wise dilation, and point-wise convolution layers. Despite LKA's efficiency in enabling vast receptive filed, we notice that the dilation convolution in LKA may cause blocking artifacts, which hurt the restored performance. Additionally, a fixed-size LKA is inflexible to fully exploit the pending features since surrounding and remote pixels play equal roles in reconstruction.



Motivated by these issues, we propose multi-scale large kernel attention (MLKA) that combines classical multi-scale mechanism and emerging LKA to build various-range correlations with relatively few computations. The multi-scale kernel can implicitly encode features from coarse to fine, which allows the model to mimic both CNNs and transformers. Moreover, to avoid potential block artifacts aroused by dilation, we adopt the gate mechanism to recalibrate the generated attention maps adaptively. To maximize the benefits of MLKA, we place it on the MetaFormer~\cite{MetaFormer}-style (\emph{Norm-TokenMixer-Norm-MLP}) structure rather than RCAN-style (\emph{Conv-Act-Conv-TokenMixer}) to construct a multi-attention block (MAB). Although transformer-style MAB can deliver higher performance, the MLP feed-forward module is too heavy for large images. Inspired by recent work~\cite{NAFNet,PvTv2}, we propose a simplified gated spatial attention unit (GSAU) by applying spatial attention and gate mechanism to reduce calculations and include spatial information. Arming with the simple yet striking MLKA and GSAU, the MABs are stacked to build the multi-scale attention network (MAN) for the SR task. In \cref{fig:Network_Performance}, we present the superior performance of our MAN. To summarize, our contributions are as follows:
\begin{itemize}
    \item We propose multi-scale large kernel attention (MLKA) for obtaining long-range dependencies at various granularity levels by combining large kernel with gate and multi-scale mechanisms, which significantly increases model representation capability.
    \item We integrate gate mechanisms and spatial attention to construct a simplified feed-forward network called GSAU which has better performance than a multi-layer perceptron (MLP) while reducing parameters and calculations.
    \item Through simply stacking the proposed modules, we present multi-scale attention network (MAN) family capable of achieving a trade-off between model complexity and performance in both lightweight and performance-oriented SR tasks.
\end{itemize}

\section{Related Work}

\subsection{Single Image Super-Resolution}
Numerous deep-learning models~\cite{VDSR,FSRCNN,EFDN,SeemoRe} have been proposed for SISR since the pioneering work SRCNN~\cite{SRCNN} introduced a 3-layer convolutional neural network (CNN) to map the correlation between LR and HR images. Depending on the model complexity, we can classify these solutions as the classical (performance-oriented) SR and the lightweight SR. 

For the classical SR task, models are delicately designed for better reconstruction quality. Specifically, VDSR~\cite{VDSR} and EDSR~\cite{EDSR} were proposed to exploit deeper information by residual learning and increasing depth and width. RCAN~\cite{RCAN} then developed EDSR by introducing channel attention (CA) and residual in residual (RIR) to further excavate intermediate features. After RCAN, many works~\cite{SAN,HAN,ENLCA} added attention mechanisms to the EDSR structure to boost performance. Recently, vision transformers~\cite{IPT,SwinIR} with self-attention (SA) were introduced to improve image restoration and refresh the SOTA performance.


For tiny and lightweight SR, the model size is constrained for mobile device deployment. The recursive learning was considered effective in decreasing the parameters in DRCN~\cite{DRCN}, DRRN~\cite{DRRN}, and LapSRN~\cite{LapSRN}. However, recursively using modules only reduces model size but maintains high computation costs. More recent works leverage productive operations, \eg, channel splits and attention module, to exploit the hierarchical features. For example, IMDN~\cite{IMDN} proposed information multi-distillation and contrast-aware channel attention. 


\subsection{Attention in Super-Resolution}

The attention mechanism can be viewed as a discriminative selection process that focuses on informative regions and ignores the irrelevant noise of pending features. Many SR networks apply attention modules to exploit latent correlations among the immediate features. Following RCAN~\cite{RCAN} that first adopted channel attention, SAN~\cite{SAN} leveraged second-order channel attention to adapt the channel-wise features through second-order statistics. Several works introduced spatial attention to enrich the feature maps, \eg, enhanced spatial attention in RFANet~\cite{RFANet}, and spatial-channel attention in HAN~\cite{HAN}. Additional CNN-based works have utilized and refined non-local attention (NLA) to obtain long-range correlations \cite{NLSA,ENLCA} and achieved an appreciable performance gain. Inspired by vision transformers~\cite{Swin,PvTv2}, self-attention has been employed in SR to capture long-term adaptability, \eg, IPT~\cite{IPT} and SwinIR~\cite{SwinIR}. More recently, DAT~\cite{DAT} leveraged SA along both channel and spatial dimensions and enabled an effective information aggregation to achieve a prominent record. GRL~\cite{GRL} utilized varied SA to explicitly model image hierarchies from coarse to fine to improve the recovery quality.

\section{Methodology}
\label{sec:methodology}

\subsection{Network Architecture}
\begin{figure*}[!t]
\centering
\includegraphics[width=0.9\linewidth]{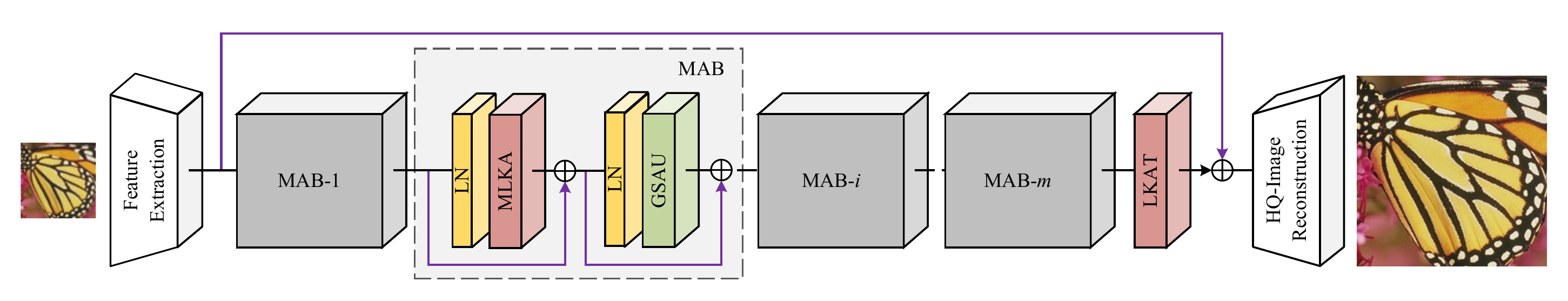} 
\caption{
Overview of our multi-scale attention network (MAN).}
\label{fig:Network_Arch}
\end{figure*}

As illustrated in \cref{fig:Network_Arch}, the proposed MAN is constituted of three components: the shallow feature extraction module (SF), the deep feature extraction module (DF) based on multiple multi-scale attention blocks (MAB), and the high-quality image reconstruction module. Given an input LR image $I_{LR}\in \mathbb{R}^{3\times H\times W}$, the SF module is first utilized to extract the primitive feature $F_{p}\in \mathbb{R}^{C\times H\times W}$ by a single 3$\times$3 convolution function $f_{SF}(\cdot)$ as follows:
\begin{equation}
    F_{p} = f_{SF}(I_{LR}).
\end{equation}

The $F_{p}$ is then sent to cascading MABs for further extraction, termed as $f_{DF}(\cdot)$, which can be formulated as:
\begin{equation}
    F_{r} = f_{DF}(F_{p}),
\end{equation}
where the $F_{r}$ is the estimated high-frequency feature for final restoration. By adding the long residual feature, the final reconstruction component restores the HQ images $I_{SR}\in \mathbb{R}^{3\times H\times W}$ by:
\begin{equation}
    I_{SR} = f_{RC}(F_{p}+F_{r}),
\end{equation}
where $f_{RC}(\cdot)$ represents reconstruction module implemented by a 3$\times$3 convolution and a pixel-shuffle layer for efficiency.

In terms of optimization, we utilize the widely used $\ell_1$ loss for a fair comparison with state-of-the-art methods~\cite{EDSR,RCAN,HAN}. Specifically, supposing an input batch of $N$ images, \ie $\{I^{LR}_i,I^{HR}_i\}^N_{i=1}$, training MAN is to minimize the $\ell_1$: 
\begin{equation}
     \ell_1(\Theta) = \frac{1}{N}\sum_{i=1}^N\left\| f_{MAN}(I^{LR}_i)-I^{HR}_i\right\|_1    
\end{equation}
where $f_{MAN}(\cdot)$ is the proposed network and $\Theta$ denotes its trainable parameters.

\subsection{Multi-scale Attention Block (MAB)}
\begin{figure}[!t]
\centering
\includegraphics[width=1.0\linewidth]{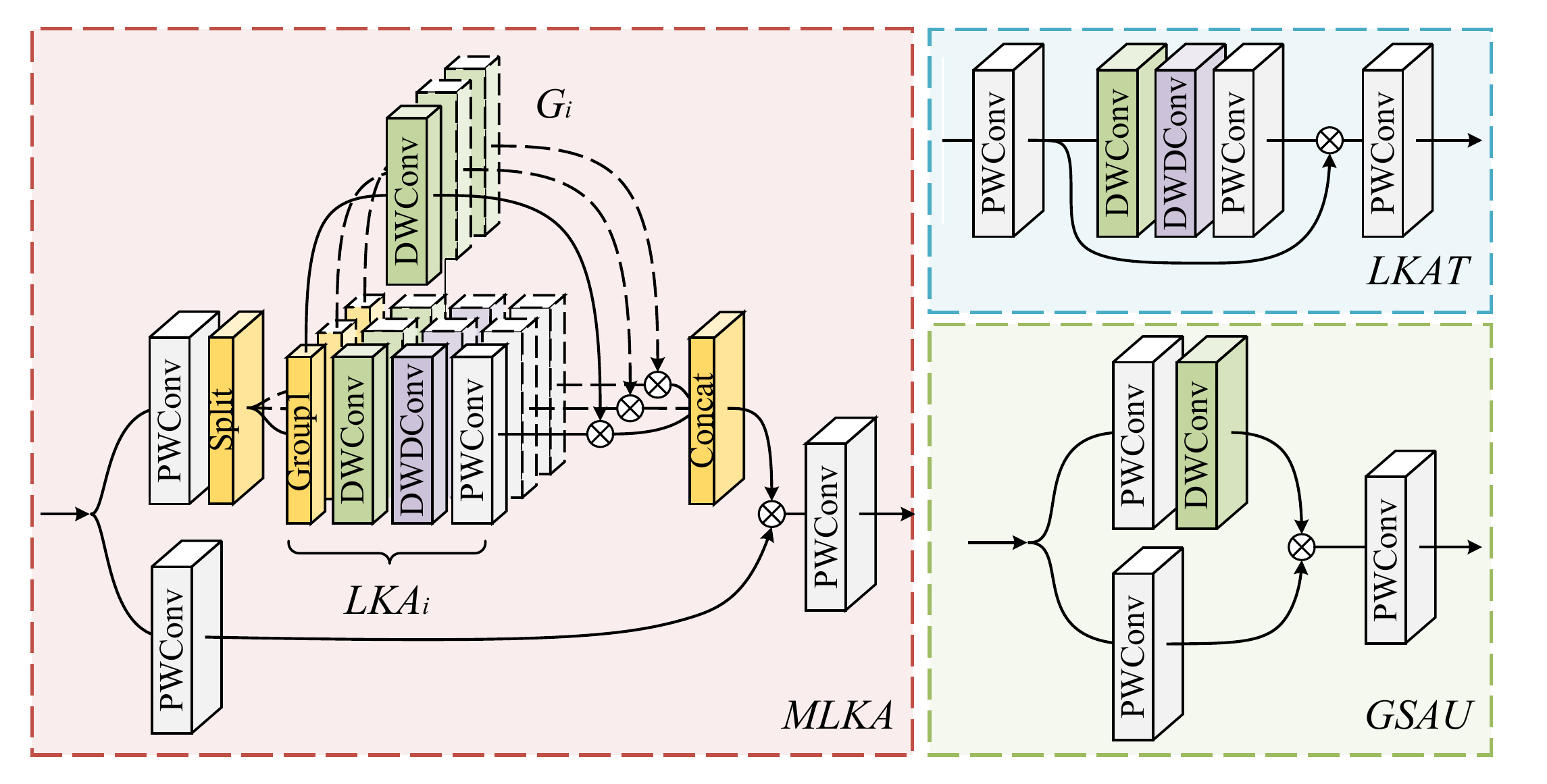} 
\caption{
Details of proposed modules.}
\label{fig:Network_Detail}
\end{figure}

Inspired by recent breakthroughs in transformers, we reconsider the basic convolutional block for feature extraction in the SISR task. In contrast to many RCAN~\cite{RCAN}-style blocks, the proposed block incorporates MetaFormer~\cite{MetaFormer}-style functionality to achieve a promising extraction result. As shown in \cref{fig:Network_Detail}, MAB consists of two components: the multi-scale large kernel attention (MLKA) module and the gate spatial attention unit (GSAU).

Given input feature $X$, the whole process of MAB is:
\begin{equation}
\begin{aligned}
   && N &= LN(X),\\
   && X &= X + \lambda_1 f_3(MLKA(f_1(N))\otimes f_2(N)), \\
   && N &= LN(X),\\
   && X &= X + \lambda_2 f_6(GSAU(f_4(N),f_5(N))),    
\end{aligned}
\end{equation}
where $LN(\cdot)$ and $\lambda$ are layer normalization and learnable scaling factors, separately. $MLKA(\cdot)$ and $GSAU(\cdot)$ are proposed MLKA and GSAU modules introduced in the following sections. $\otimes$ and $f_i(\cdot)$ represent element-wise multiplication and $i$-th point-wise convolution that keeps the dimensions. To preserve instance details and accelerate convergence, we employ layer normalization rather than batch normalization or none normalization.

\subsection{Multi-scale Large Kernel Attention (MLKA)} \label{subsec:MLKA}
The attention mechanism can force networks to focus on crucial information and ignore irrelevant ones. Previous SR models adopt a series of attention mechanisms, including channel attention~(CA) and self-attention~(SA), to obtain more informative features. However, these methods fail to simultaneously uptake local information and long-range dependence, and they often consider the attention maps at a fixed reception field. Enlightened by the latest visual attention research~\cite {VAN}, we propose multi-scale large kernel attention (MLKA) to resolve these problems by combining large kernel decomposition and multi-scale learning. Specifically, the MLKA consists of three main functions, large kernel attentions~(LKA) for establishing interdependence, the multi-scale mechanism for obtaining heterogeneous-scale correlation, and gated aggregation for dynamic recalibration.

\textbf{Large kernel attention.} Given the input feature maps $X\in \mathbb{R}^{C\times H\times W}$, the LKA adaptively builds the long-range relationship by decomposing a $K\times K$ convolution into three components: a $(2d-1)\times (2d-1)$ depth-wise convolution $f_{DW}(\cdot)$, a $\lceil\frac{K}{d}\times \frac{K}{d}\rceil$ depth-wise $d$-dilation convolution $f_{DWD}(\cdot)$, and a point-wise convolution $f_{PW}(\cdot)$, which can be formulated as:  
\begin{equation}
    LKA(X) = f_{PW}(f_{DWD}(f_{DW}(X))).
\end{equation}

\textbf{Multi-scale mechanism.} To learn the attention maps with omni-scale information, we enhance the fixed LKA with the group-wise multi-scale mechanism. Supposing the input feature maps $X\in \mathbb{R}^{C\times H\times W}$, the module first splits it into $n$-pieces $X_1,X_2,\dots,X_n$ of ${\lfloor\frac{C}{n}\rfloor\times H\times W}$. For $i$-th group of features $X_i$, an LKA decomposed by $\{K_i,d_i\}$ is utilized to generate a homogeneous scale attention map $LKA_i$. In detail, we leverage three groups of LKA: $\{7, 2\}$ implemented by 3-5-1, $\{21, 3\}$ by 5-7-1, and $\{35, 4\}$ by 7-9-1, where $a$-$b$-1 means cascading $a\times a$ depth-wise, $b\times b$ depth-wise-dilated, and point-wise convolutions.           

\begin{figure}[!t]
\setlength\tabcolsep{0.0pt}
\fontsize{8.5pt}{9.5pt}\selectfont
\centering
\begin{tabular}{ccc}
\includegraphics[height=4.2cm]{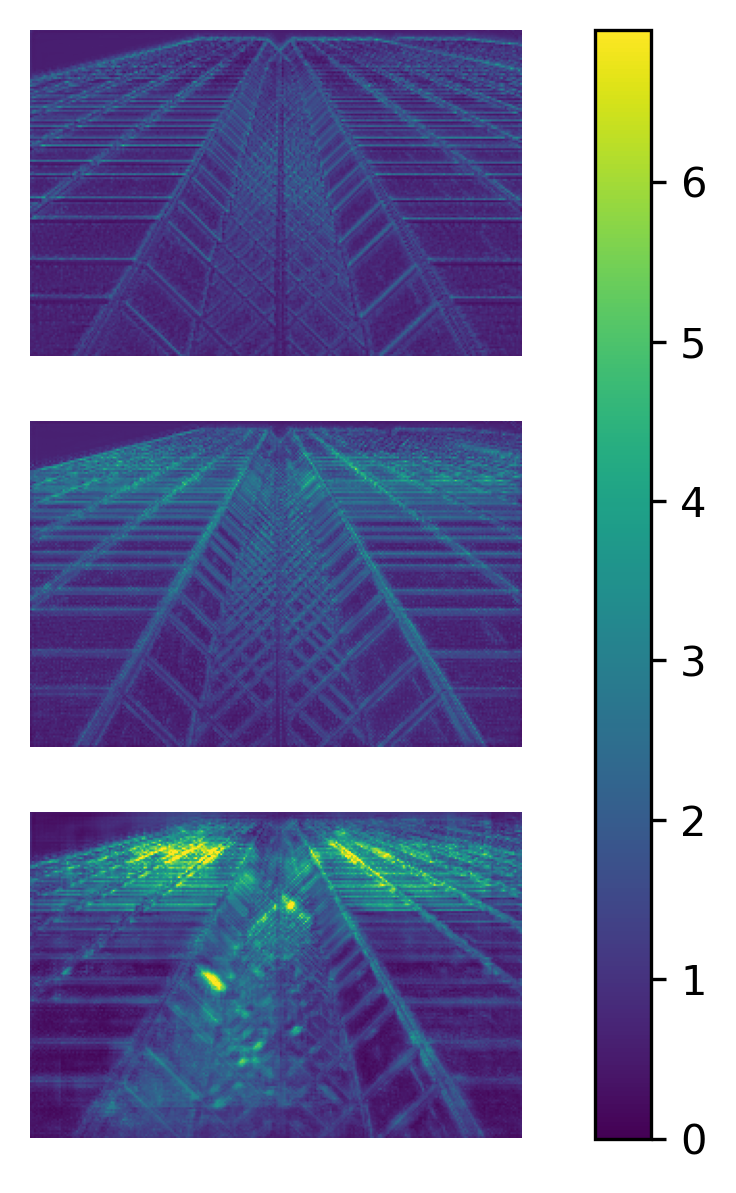} 
& \includegraphics[height=4.2cm]{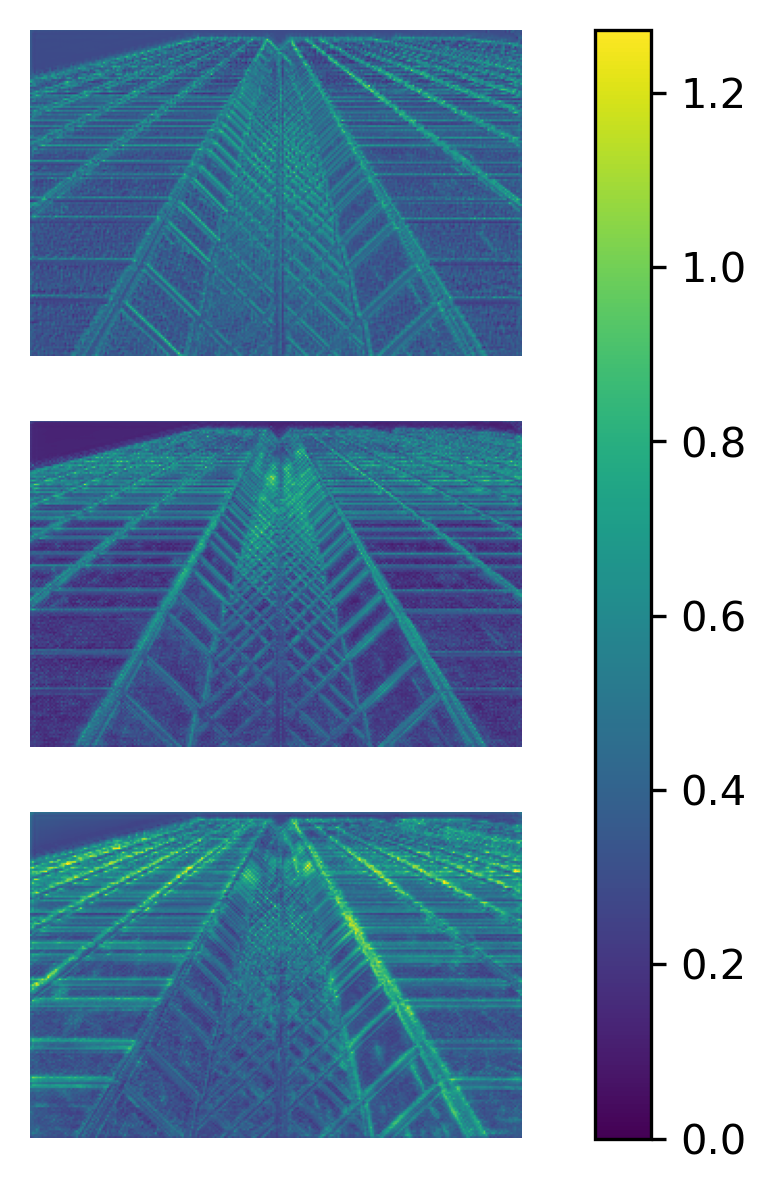}
& \includegraphics[height=4.2cm]{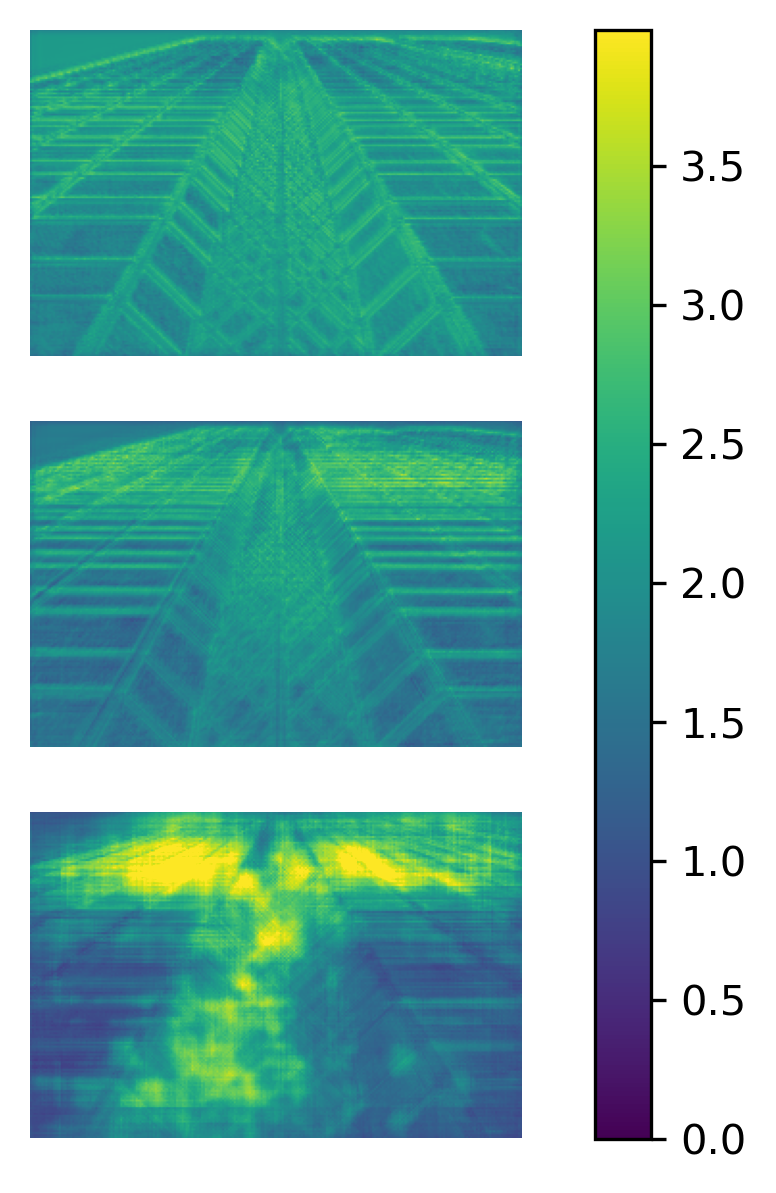} \\
$MLKA_i\quad$& $G_i\quad$  & $LKA_i\quad$\\
\end{tabular}
\caption{Visual activation maps of \cref{eq:GSAU} in the 16-th layer of MAN-light. From top to bottom are the corresponding feature maps of 3-5-1, 5-7-1, and 7-9-1, respectively. }
\label{fig:AG}
\end{figure}

\textbf{Gated aggregation.} Different from many high-level computer vision tasks, the SR task has a worse tolerance for dilation and partition. As shown in the \cref{fig:AG}, although the larger LKA captures wider responses of pixels, the blocking artifacts  appear in the generated attention maps of larger LKA. For $i$-th group input $X_i$, to avoid the block effect, as well as to learn more local information, we leverage spatial gate to dynamically adapt $LKA_i(\cdot)$ into $MLKA_i(\cdot)$ by:
\begin{equation}
    MLKA_i(X_i) = G_i(X_i)\otimes LKA_i(X_i),
\end{equation}
where $G_i(\cdot)$ is the $i$-th gate generated by $a_i\times a_i$ depth-wise convolution, and $LKA_i(\cdot)$ is the LKA decomposed by $a_i$-$b_i$-1. In \cref{fig:AG}, we provide the visual results of the gated aggregation. It can be observed that the block effects are removed from the attention maps and the $MLKA_i$s are more reasonable. In particular, the $MLKA_i$ with larger receptive fields reacts more on long-range dependence while the smaller $MLKA_i$ tends to retain local texture. 

\textbf{Complexity analysis.} To compare the complexities of MLKA, LKA, and SA, we present their theoretical floating point operation (FLOPs). Given input $X\in \mathbb{R}^{C\times H\times W}$, the computational cost of $M\times M$ window-based SA is $2M^2HWC$. Within LKA with fixed $\{K,d\}$, the budget of decomposition is ${(\lceil\frac{K}{d}\rceil^2+(2d-1)^2+C)HWC}$. In general, the window size $M$ and kernel size $K$ determine theoretical computational complexity by the quadratic increase. For the proposed MLKA with $n$ groups of $\{K_i,d_i\}$, the total computation is denoted as $(\sum{\frac{1}{n}(\lceil\frac{K_i}{d_i}\rceil^2+{\color{blue}2}(2d_i-1)^2+\frac{C}{n})+{\color{blue}C})HWC}$. The {\color{blue}blue} terms are the additional calculations brought by gated aggregation and projection. Since the feature is separated into small groups (divided by $n$), we can control the computational cost while flexibly employing varied kernels to capture both local and global information.

\subsection{Gated Spatial Attention Unit (GSAU)} \label{subsec:GSAU}
In transformer blocks, a feed-forward network (FFN) is an essential part of enhancing feature representation. However, the commonly used MLP with wide intermediate channels is too heavy for SR networks, especially for large image inputs. Inspired by \cite{GLU,NAFNet,FLASH,PvTv2}, we integrate simple spatial attention (SSA) and gated linear unit (GLU) into the proposed GSAU to enable an adaptive gating mechanism and reduce the parameters and calculations. 

To capture spatial information more efficiently, we adopt a single layer depth-wise convolution to weight the feature map. Given the dense-transformed $X$ and $Y$, the key process of GSAU can be represented as:
\begin{equation}
    GSAU(X,Y) = f_{DW}(X)\otimes Y,
    \label{eq:GSAU}
\end{equation} 
where $f_{DW}(\cdot)$ and $\otimes$ indicate depth-wise convolution and element-wise multiplication, respectively. By applying a spatial gate, the GSAU can remove the nonlinear layer and capture local continuity under considerate complexity.

\begin{table*}[!t] 
\center
\begin{center}
\caption{Ablation studies on components of MAN. The impact of LKAT, Multi-scale mechanism, and GSAU are shown upon MAN-tiny/light ($\times 2$). In detail, we replace LKAT with convolution layer, Multi-scale with LKA~(5-7-1), and GSAU with MLP.}
\label{tab:ablation_study1}
\small
\fontsize{8.5pt}{9.5pt}\selectfont
\tabcolsep=1.8pt
\begin{tabular}{l|c|c|c|c|c|c|c|c|c|c|c|c|c}
\whline
\multirow{2}{*}{Method} & \multirow{2}{*}{LKAT} &
\multirow{2}{*}{\makecell[c]{Multi-\\Scale}} &\multirow{2}{*}{GSAU} &
\multirow{2}{*}{\#Params} & \multirow{2}{*}{\#Mult-Adds} &  \multicolumn{2}{c|}{Set5~\cite{Set5}} &  \multicolumn{2}{c|}{Set14~\cite{Set14}} &  \multicolumn{2}{c|}{BSD100~\cite{B100}} &  \multicolumn{2}{c}{Urban100~\cite{Urban100}}
\\
\cline{7-14}
&  &  &  & & & PSNR\,{\scriptsize ($\Delta$)} & SSIM & PSNR\,{\scriptsize ($\Delta$)} & SSIM & PSNR\,{\scriptsize ($\Delta$)} & SSIM & PSNR\,{\scriptsize ($\Delta$)} & SSIM 
\\
\whline
\multirow{4}{*}{\algname{}-tiny} \qquad  & & & & 108K & {24.2G} %
& {37.71}\,{\scriptsize\color{white}(\textuparrow 0.00)}
& {0.9594}
& {33.24}\,{\scriptsize\color{white}(\textuparrow 0.00)}
& {0.9148}
& {31.97}\,{\scriptsize\color{white}(\textuparrow 0.00)}
& {0.8973}
& {31.08}\,{\scriptsize\color{white}(\textuparrow 0.00)}
& {0.9178}
\\
  \qquad\qquad  & \checkmark & & & 121K & {27.2G} %
& {37.75}\,{\scriptsize\color{green}(\textuparrow 0.04)}
& {0.9595}
& {33.27}\,{\scriptsize\color{green}(\textuparrow 0.03)}
& {0.9154}
& {32.01}\,{\scriptsize\color{green}(\textuparrow 0.04)}
& {0.8979}
& {31.25}\,{\scriptsize\color{green}(\textuparrow 0.17)}
& {0.9199}
\\
  & \checkmark & \checkmark & & 143K & {29.9G} %
& {37.77}\,{\scriptsize\color{green}(\textuparrow 0.06)}
& {0.9596}
& {33.30}\,{\scriptsize\color{green}(\textuparrow 0.06)}
& {0.9153}
& {32.01}\,{\scriptsize\color{green}(\textuparrow 0.04)}
& {0.8979}
& {31.30}\,{\scriptsize\color{green}(\textuparrow 0.22)}
& {0.9202}
\\
  & \checkmark & \checkmark & \checkmark & 134K & {29.9G} %
& {37.79}\,{\scriptsize\color{green}(\textuparrow 0.08)}
& {0.9598}
& {33.31}\,{\scriptsize\color{green}(\textuparrow 0.07)}
& {0.9155}
& {32.02}\,{\scriptsize\color{green}(\textuparrow 0.05)}
& {0.8980}
& {31.33}\,{\scriptsize\color{green}(\textuparrow 0.25)}
& {0.9206}
\\
\whline
\multirow{4}{*}{\algname{}-light} \qquad & & & & 737K & {165.8G} %
& {38.01}\,{\scriptsize\color{white}(\textuparrow 0.00)}
& {0.9605}
& {33.55}\,{\scriptsize\color{white}(\textuparrow 0.00)}
& {0.9179}
& {32.23}\,{\scriptsize\color{white}(\textuparrow 0.00)}
& {0.9005}
& {32.14}\,{\scriptsize\color{white}(\textuparrow 0.00)}
& {0.9287}
\\
  \qquad\qquad  & \checkmark & & & 756K & {170.0G} %
& {38.05}\,{\scriptsize\color{green}(\textuparrow 0.04)}
& {0.9607}
& {33.60}\,{\scriptsize\color{green}(\textuparrow 0.05)}
& {0.9182}
& {32.25}\,{\scriptsize\color{green}(\textuparrow 0.02)}
& {0.9007}
& {32.23}\,{\scriptsize\color{green}(\textuparrow 0.09)}
& {0.9297}
\\
  & \checkmark & \checkmark & & 835K & {187.6G} %
& {38.07}\,{\scriptsize\color{green}(\textuparrow 0.06)}
& {0.9607}
& {33.62}\,{\scriptsize\color{green}(\textuparrow 0.07)}
& {0.9181}
& {32.26}\,{\scriptsize\color{green}(\textuparrow 0.03)}
& {0.9009}
& {32.42}\,{\scriptsize\color{green}(\textuparrow 0.28)}
& {0.9308}
\\
  & \checkmark & \checkmark & \checkmark & 820K & {184.0G} %
& {38.07}\,{\scriptsize\color{green}(\textuparrow 0.06)}
& {0.9608}
& {33.69}\,{\scriptsize\color{green}(\textuparrow 0.14)}
& {0.9188}
& {32.29}\,{\scriptsize\color{green}(\textuparrow 0.06)}
& {0.9012}
& {32.43}\,{\scriptsize\color{green}(\textuparrow 0.29)}
& {0.9316}
\\
\whline   

\end{tabular}
\end{center}
\end{table*}

\subsection{Large Kernel Attention Tail (LKAT)}
In previous SR networks~\cite{EDSR,RCAN,SAN,HAN,SwinIR}, the vanilla convolution layer is widely used as the tail of the deep extraction backbone. However, it has a flaw in establishing long-range connections, therefore limiting the representative capability of the finial reconstruction feature. In order to summarize more reasonable information from the stacked MABs, we introduce the 7-9-1 LKA in the tail module. Concretely, the LKA is wrapped by two $1\times1$ convolutions as depicted in \cref{fig:Network_Detail}. 

\section{Experiments}
\subsection{Datasets and Metrics}
Following latest works \cite{DFSA,SwinIR,RCAN_it}, we utilize DIV2K~\cite{div2k} and Flickr2K~\cite{EDSR}, which contain 800 and 2650 images, to train our models. For the testing phase, we evaluate our method on five commonly used datasets: Set5~\cite{Set5}, Set14~\cite{Set14}, BSD100~\cite{B100}, Urban100~\cite{Urban100}, and Manga109~\cite{Manga109}. In addition, two standard evaluation metrics, peak-signal-to-noise-ratio (PSNR) and the structural similarity index (SSIM)~\cite{SSIM}, are applied in $Y$ channel of the YCbCr images to measure the quality of restoration.

\subsection{Implementation  Details}
For more comprehensive evaluations of the proposed methods, we train three different versions of MAN: tiny, light, and base, to resolve the classic SR tasks under different complexities. Following \cite{SwinIR}, we stack 5/24/36 MABs and set the channel width to 48/60/180 in the corresponding tiny/light/base MAN. Three multi-scale decomposition modes are utilized in MLKA, listed as 3-5-1, 5-7-1, and 7-9-1. The $7\times 7$ depth-wise convolution is used in the GSAU. 

In the training stage, the training pairs are augmented by
horizontal flips and random rotations of 90\textdegree, 180\textdegree, and 270\textdegree. The \{patch size, batch size\} is set to \{$48\times 48$, 32\} and \{$64\times 64$, 16\} in the training-from-scratch and fine-turning stage, respectively. The $\ell_1$ loss is adopted to discriminate the pixel-wise restoration quality for fairness. All models are trained using the Adam optimizer~\cite{ADAM} with $\beta_1$=0.9, $\beta_2$=0.99. The learning rate is initialized as 5$\times$10$^{-4}$ and scheduled by cosine annealing learning for 1600K iterations in training anew while setting as 1$\times$10$^{-4}$ for 800K in fine-turning. 
All experiments are conducted by Pytorch~\cite{Pytorch} framework on 4 Nvidia RTX 3090 GPUs.

\begin{table}[!t]\scriptsize
\center
\begin{center}
\caption{Ablation study on block structure.}

\label{tab:ablation_study2}
\small
\fontsize{8.5pt}{9.5pt}\selectfont
\tabcolsep=1pt
\begin{tabular}{l|c|c|c|c|c|c}
\whline
\multirow{2}{*}{Method} &
\multirow{2}{*}{\#Params} & \multirow{2}{*}{\#FLOPs} &  \multicolumn{2}{c|}{Set5~\cite{Set5}} &  \multicolumn{2}{c}{BSD100~\cite{B100}} 
\\
\cline{4-7}
& & & PSNR & SSIM & PSNR & SSIM 
\\
\whline
RCAN-style & 924K & {53.0G} %
& {32.16}
& {0.8945}
& {27.60}
& {0.7371}
\\
\rowcolor{tableblue}
Metaformer-style & 840K & {47.1G} %
& {32.33}
& {0.8967}
& {27.67}
& {0.7396}
\\
\whline
\end{tabular}
\end{center}
\end{table}

\begin{table*}[htb]

\caption{Ablation studies of multi-scale and decomposition type (LKA/MLKA). 
The results are tested on MAN-light for $\times$4 SR task. The LKA~(5-7-1) from VAN~\cite{VAN} is used as the baseline for comparison.  } 

\begin{minipage}[p]{0.73\textwidth}
\label{tab:ablation_study3}
\centering
\small
\fontsize{8.5pt}{9.5pt}\selectfont
\tabcolsep=1pt
\begin{tabular}{l|c|c|c|c|c|c|c|c|c}
\whline
\multirow{2}{*}{Method} & \multicolumn{3}{c|}{Decomposition} &
\multirow{2}{*}{\#Params} & \multirow{2}{*}{\#FLOPs} &  \multicolumn{2}{c|}{Set5~\cite{Set5}} &  \multicolumn{2}{c}{Set14~\cite{Set14}}    
\\
\cline{7-10}\cline{2-4}
&3-5-1  &5-7-1  &7-9-1  & & & PSNR\,{\scriptsize ($\Delta$)} & SSIM\,{\scriptsize ($\Delta$)} & PSNR\,{\scriptsize ($\Delta$)} & SSIM\,{\scriptsize ($\Delta$)}
\\
\whline
\multirow{3}{*}{LKA} & \checkmark & & & 703K & {39.4G} %
& {32.23}\,{\scriptsize\color{red}(\textdownarrow
 0.04)}
& {0.8956}\,{\scriptsize\color{red}(\textdownarrow
 .0007)}
& {28.70}\,{\scriptsize\color{red}(\textdownarrow
 0.02)}
& {0.7842}\,{\scriptsize\color{red}(\textdownarrow
 .0004)}
\\
  &  & \checkmark & & 761K & {42.7G} %
& {32.27}\,{\scriptsize\color{white}(\textuparrow 0.00)}
& {0.8963}\,{\scriptsize\color{white}(\textuparrow .0000)}
& {28.72}\,{\scriptsize\color{white}(\textuparrow 0.00)}
& {0.7846}\,{\scriptsize\color{white}(\textuparrow .0000)}
\\
  & & & \checkmark & 841K & {47.4G} %
& {32.25}\,{\scriptsize\color{red}(\textdownarrow
 0.02)}
& {0.8958}\,{\scriptsize\color{red}(\textdownarrow
 .0005)}
& {28.71}\,{\scriptsize\color{red}(\textdownarrow
 0.01)}
& {0.7845}\,{\scriptsize\color{red}(\textdownarrow
 .0001)}
\\
\hdashline
\multirow{3}{*}{MLKA} & \checkmark & \checkmark & & 803K & {45.0G} %
& {32.32}\,{\scriptsize\color{green}(\textuparrow 0.05)}
& {0.8968}\,{\scriptsize\color{green}(\textuparrow.0005)}
& {28.72}\,{\scriptsize\color{green}(\textuparrow 0.00)}
& {0.7848}\,{\scriptsize\color{green}(\textuparrow.0002)}
\\
  & & \checkmark & \checkmark & 900K & {50.6G} %
& {32.33}\,{\scriptsize\color{green}(\textuparrow 0.06)}
& {0.8968}\,{\scriptsize\color{green}(\textuparrow.0005)}
& {28.74}\,{\scriptsize\color{green}(\textuparrow 0.02)}
& {0.7852}\,{\scriptsize\color{green}(\textuparrow.0006)}
\\
  & \checkmark & \checkmark & \checkmark & 840K & {47.1G} %
& {32.33}\,{\scriptsize\color{green}(\textuparrow 0.06)}
& {0.8967}\,{\scriptsize\color{green}(\textuparrow.0004)}
& {28.76}\,{\scriptsize\color{green}(\textuparrow 0.04)}
& {0.7856}\,{\scriptsize\color{green}(\textuparrow.0010)}
\\
\whline   
\end{tabular}

\centerline{\qquad}\leavevmode\\
\end{minipage}
\begin{minipage}[p]{0.283\textwidth} 
    \centering 
    \resizebox{\linewidth}{!}{%
    \includegraphics[width=1\textwidth]{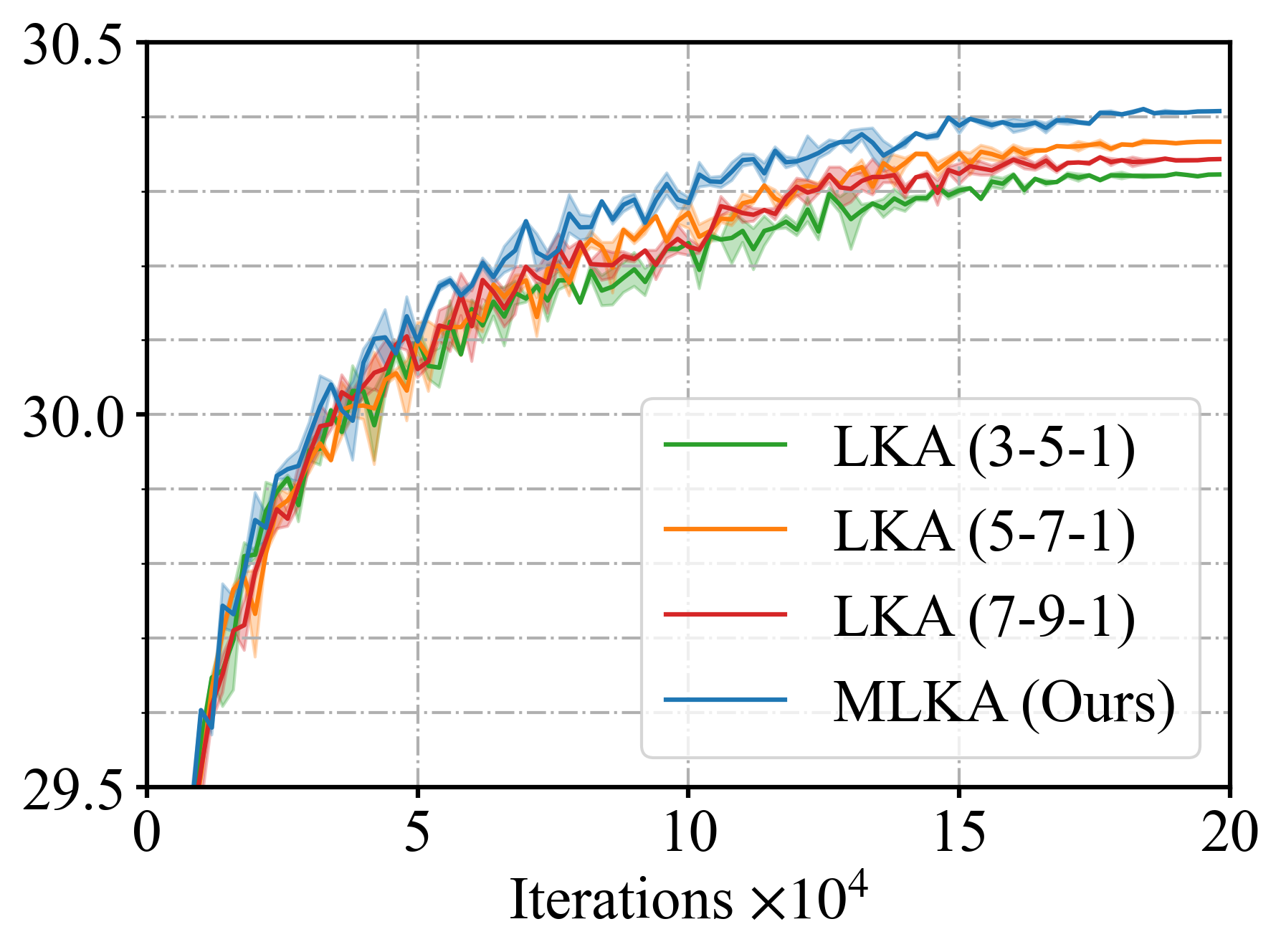} 
    }
\end{minipage}
\vspace{-3mm}
\label{tab1:env} 
\end{table*}

\begin{table}[!t]
\center
\begin{center}
\vspace{-5mm}
\caption{Ablation study on varied FFNs.}
\label{tab:ablation_study4}
\small
\fontsize{8.5pt}{9.5pt}\selectfont
\tabcolsep=2pt
\begin{tabular}{l|c|c|c|c|c|c}
\whline
{Method} &
{\#Params} & {\#FLOPs} &  {Set5} & {Set14}&  {B100} & {U100}   
\\
\whline
MLP~\cite{ViT} & 854K & {48.0G} %
& {32.31}
& {28.73}
& {27.65}
& {26.26}
\\
Simple-Gate~\cite{NAFNet} & 768K & {43.1G} %
& {32.28}
& {28.74}
& {27.66}
& {26.28}
\\
CFF~\cite{PvTv2} & 1140K & {64.3G} %
& {32.35}
& {28.76}
& {27.67}
& {26.34}
\\
\rowcolor{tableblue}
GSAU \qquad\qquad\quad & 840K & {47.1G} %
& {32.33}
& {28.76}
& {27.67}
& {26.31}
\\
\whline
\end{tabular}
\end{center}
\end{table}

\subsection{Ablation Studies}
In this section, we validate the effectiveness of the proposed components from coarse to fine. In detail, we first investigate the combination of all proposed modules and then examine each of them individually. \emph{For fairness and simplicity, we adopt the same training for 200K iterations}. 

\textbf{Overall study on components of MAN.} In \cref{tab:ablation_study1}, we present the results of deploying the proposed components on our tiny and light networks. In general, the best performances are achieved by employing all proposed modules. Specifically, 0.25\,dB and 0.29\,dB promoting on Urban100~\cite{Urban100} can be observed in MAN-tiny and MAN-light, while the parameters and calculations increase negligibly. Among these components, the LKAT module and multi-scale mechanism are more important to enhance quality. Without any of them, the PSNR will drop by 0.09\,dB. The GSAU is an economical replacement for MLP. It reduces 15K parameters and 3.6G calculations while bringing significant improvements across all datasets.

\textbf{Study on block structures.} Within MAB, we choose the emerging metaformer style rather than the RCAN-style structure to deploy MLKA. To fully explore their effectiveness, we implement and compare two versions of MABs in the \cref{tab:ablation_study2}. The experimental results indicate that the transformer-style MAB surpasses the RCAN-style one by a large margin. On Set5~\cite{Set5}, the PSNR is increased from 32.15\,dB to 32.33\,dB by employing the transformer structure. The results show the transformer-style MAB is more efficient in balancing the performance and computations.

\begin{figure}[!t]
\setlength\tabcolsep{1.0pt}
\fontsize{8.5pt}{9.5pt}\selectfont
\centering
\begin{tabular}{cccccc}
&HR& LKA (3-5-1)  &LKA (5-7-1) & LKA (7-9-1) &MLKA\\
{\rotatebox{90}{\emph{img\_067}}}
&\includegraphics[width=.18\linewidth,height=.25\linewidth]{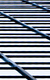} 
& \includegraphics[width=.18\linewidth,height=.25\linewidth]{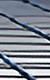} 
& \includegraphics[width=.18\linewidth,height=.25\linewidth]{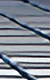} 
& \includegraphics[width=.18\linewidth,height=.25\linewidth]{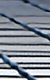} 
& \includegraphics[width=.18\linewidth,height=.25\linewidth]{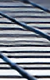} \\
{\rotatebox{90}{\emph{0884}}}& \includegraphics[width=.18\linewidth,cframe=black]{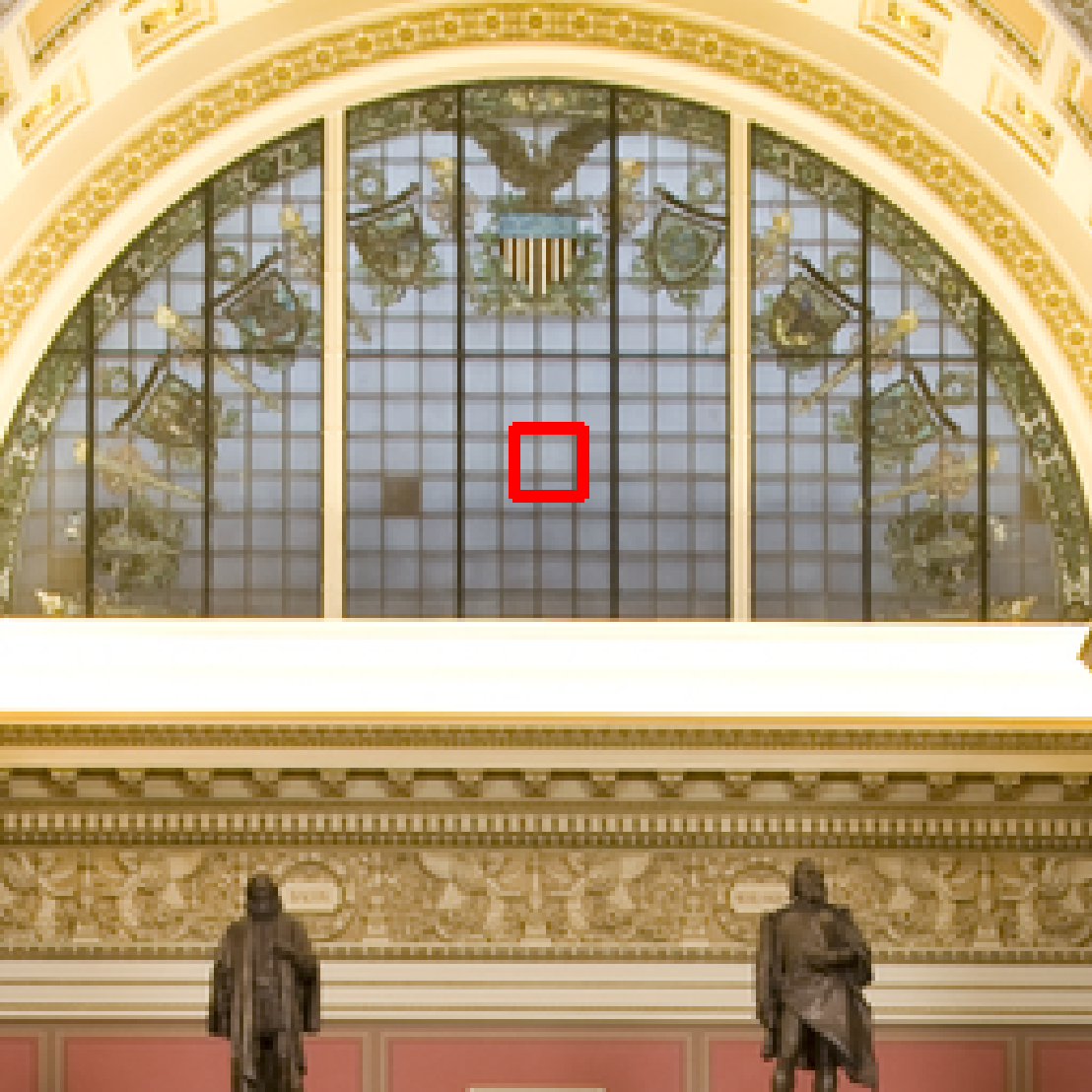}
&\includegraphics[width=.18\linewidth,cframe=black]{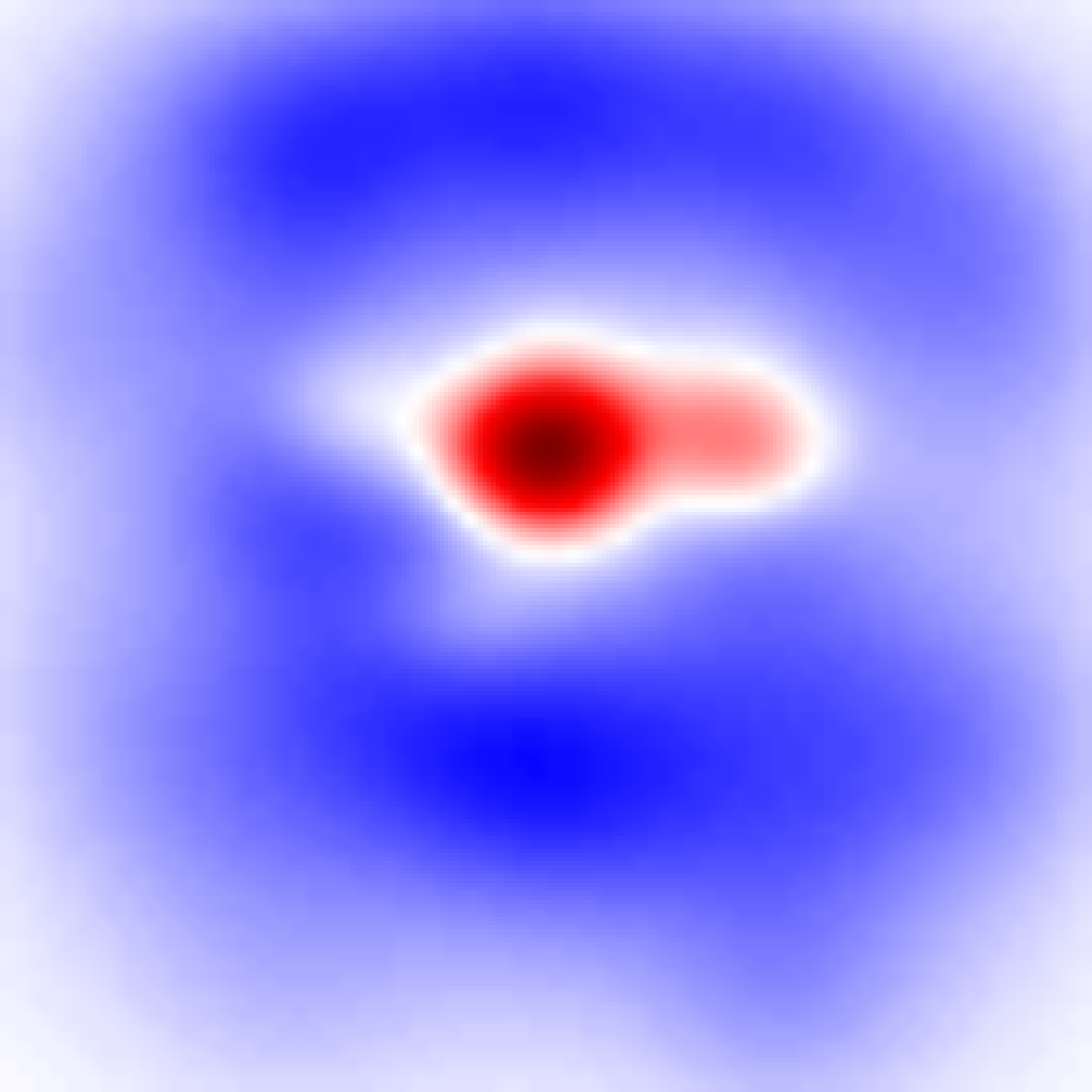}
&\includegraphics[width=.18\linewidth,cframe=black]{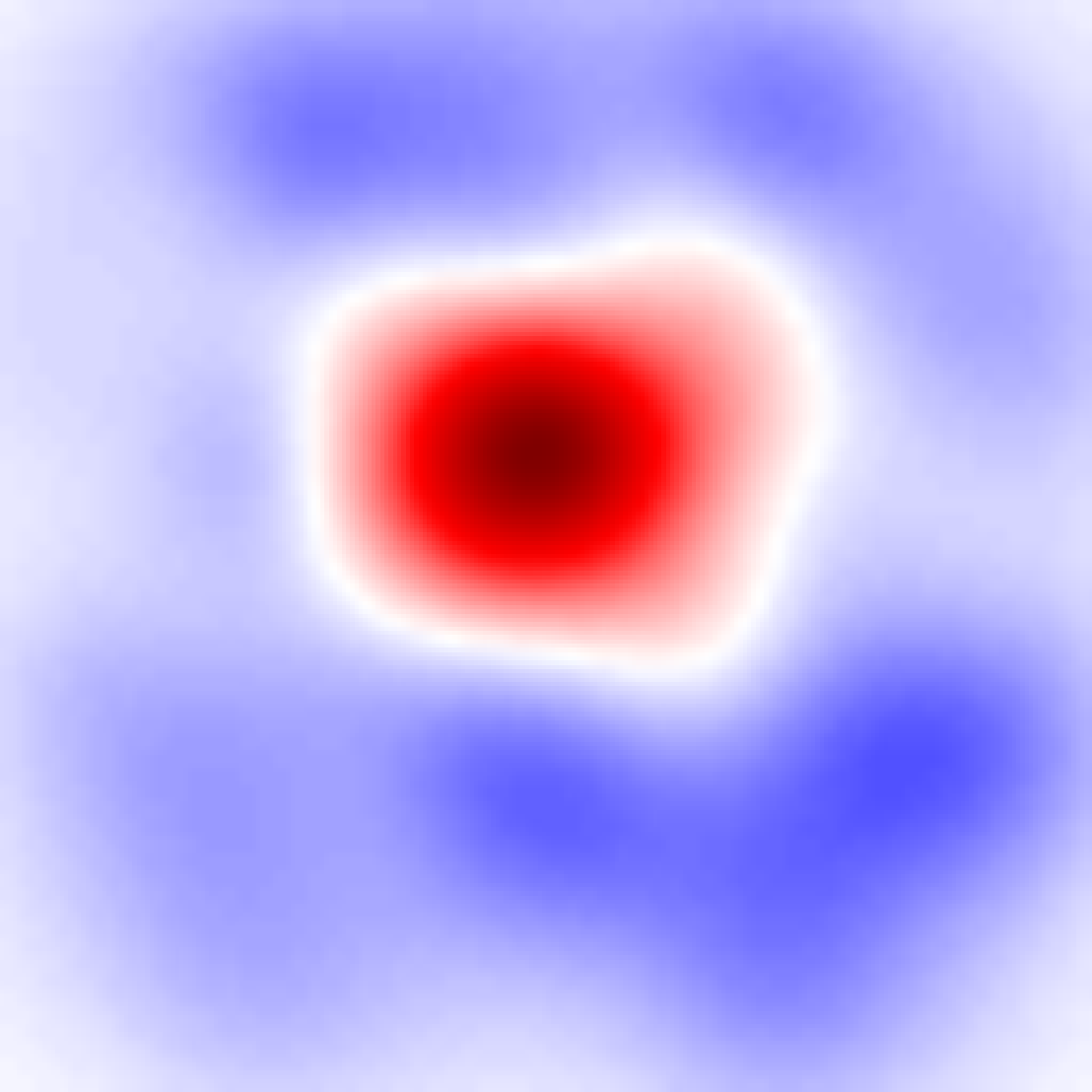}
&\includegraphics[width=.18\linewidth,cframe=black]{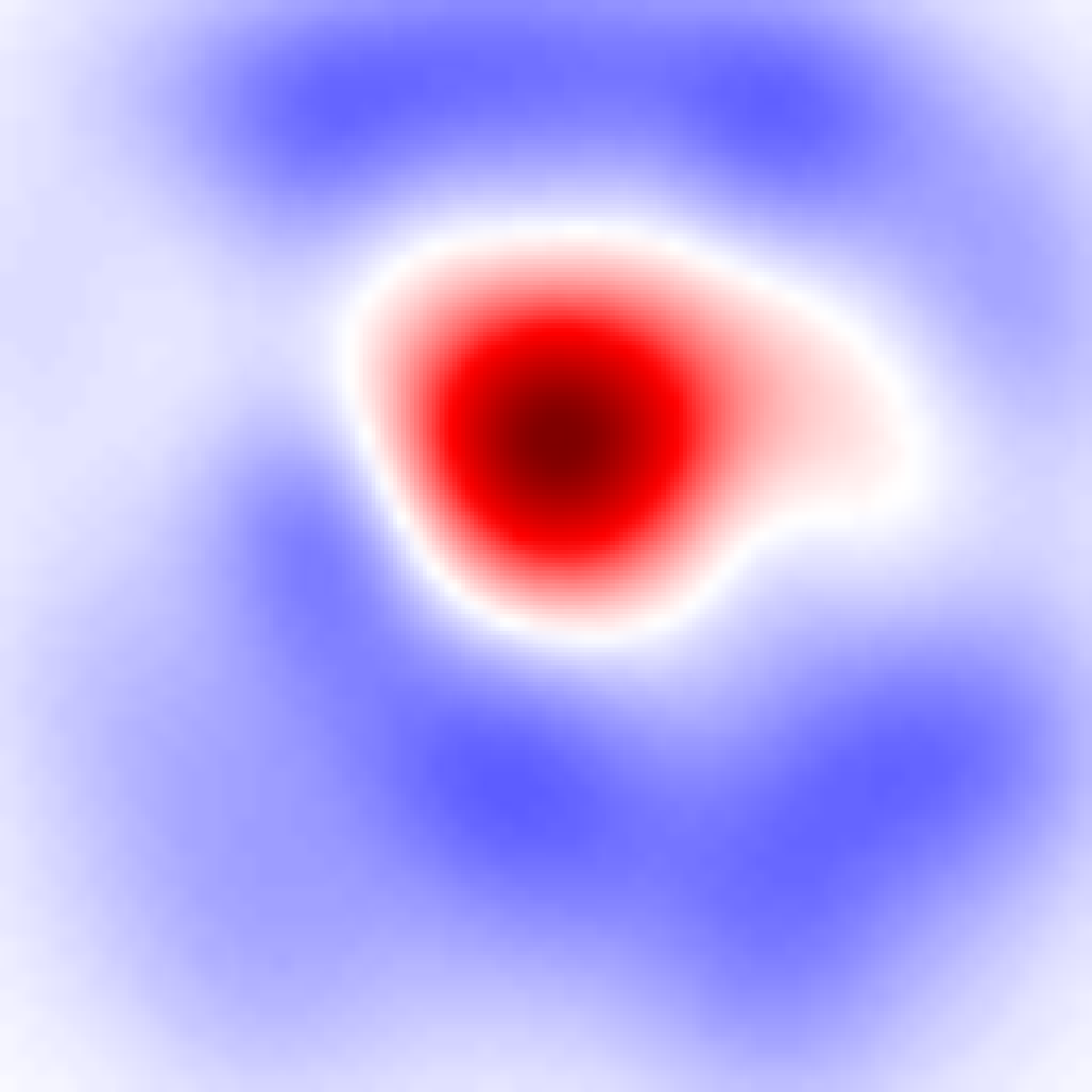}
&\includegraphics[width=.18\linewidth,cframe=black]{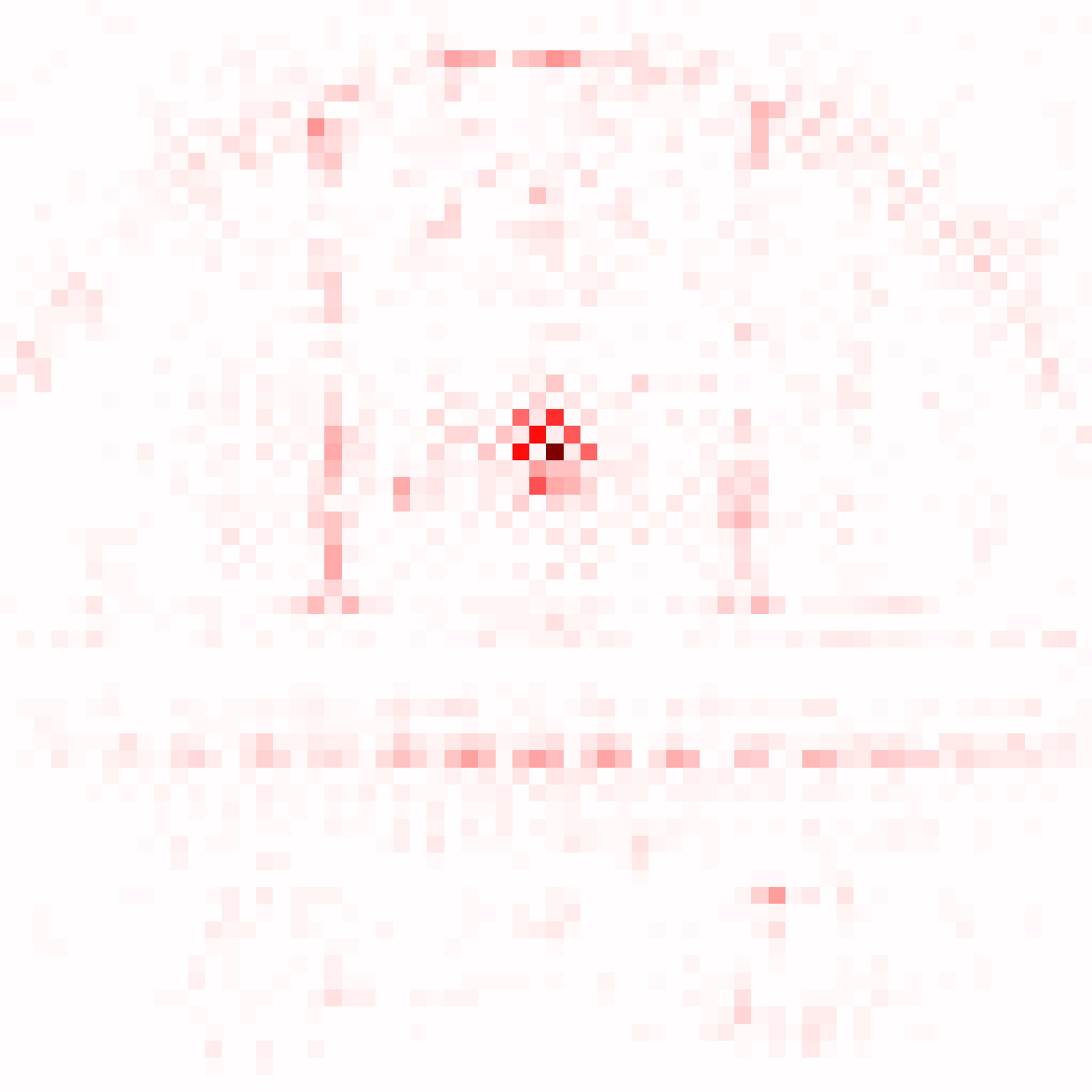}
\\
& DI & 9.5 & 27.0 & 26.4 & 32.4
\end{tabular}
\caption{Comparisons between LKA and MLKA. Rows 1: visual comparisons. Row 2: Cols 2-4: The difference maps of the area of interest between LKA and MLKA. {The \textcolor{red}{red} regions are noticed by almost both LKA and MLKA} while {the \textcolor{blue}{blue} represent additional interest areas of MLKA.} Col 5: LAM results of MLKA.}
\label{fig:2}
\end{figure}

\textbf{Study on MLKA.} To justify our design of MLKA, we conduct ablation experiments on multi-scale and kernel decomposition. Specifically, we consider three LKA and three MLKA implementations in \cref{tab:ablation_study3}. We first investigate the effects of kernel size on LKA. When we increase the kernel size, the PSNR decreases after an initial increase, which is inconsistent with high-level tasks~\cite{VAN}. It is due to long-range correlation and local textural information is indispensable in image restoration tasks. Up until this point, we introduce MLKA to refine features at comprehensive scales. In \cref{tab:ablation_study3}, we also illustrate the training evaluation results of LKAs and proposed MLKA. The MLKA outperforms other LKAs throughout the training phase. For the visual comparison and local attribution map (LAM)~\cite{LAM} results shown in \cref{fig:2}, MLKA brings higher DI and more activated pixels, thereby helping to recover more details on both images from Urban100. In addition, we briefly discuss MLKA of different combinations. These results suggest the MLKA with all three decomposition types can trade off parameters, computations, and performance.

\textbf{Study on FFNs.} To further confirm the efficiency of the proposed GSAU, we compare it with some other FFNs. In \cref{tab:ablation_study4}, we validate four advanced designs: MLP, Simple Gate, CFF, and our GSAU. The GSAU delivers comparable performance to the powerful CFF while occupying 73\% of the parameters and calculations, showing effectiveness.

\begin{table*}[!t] 
\center
\begin{center}
\caption{Quantitative comparison (average PSNR/SSIM) with state-of-the-art ConvNets for \textbf{\underline{classical image SR}}. The best and second best performances are \red{highlighted} and \blue{underlined}, respectively. ``$\dagger$'' and `+'' indicate using pre-training and self-ensemble strategy, respectively. }
\label{tab:sr_results}
\small
\fontsize{8.5pt}{9.5pt}\selectfont
\tabcolsep=4pt
\begin{tabular}{l|c|c|c|c|c|c|c|c|c|c|c|c|c}
\whline
\multirow{2}{*}{Method} & \multirow{2}{*}{Scale} & \multirow{2}{*}{\#Params} & \multirow{2}{*}{\#FLOPs} &  \multicolumn{2}{c|}{Set5~\cite{Set5}} &  \multicolumn{2}{c|}{Set14~\cite{Set14}} &  \multicolumn{2}{c|}{BSD100~\cite{B100}} &  \multicolumn{2}{c|}{Urban100~\cite{Urban100}} &  \multicolumn{2}{c}{Manga109~\cite{Manga109}}  
\\
\cline{5-14}
&  & & & PSNR & SSIM & PSNR & SSIM & PSNR & SSIM & PSNR & SSIM & PSNR & SSIM 
\\
\whline
RCAN~\cite{RCAN} & $\times$2 & 15.4M & 3.5T%
& {38.27}
& {0.9614}
& {34.12}
& {0.9216}
& {32.41}
& {0.9027}
& {33.34}
& {0.9384}
& {39.44}
& {0.9786}
\\  
SAN~\cite{SAN} & $\times$2 & 15.9M& 3.1T %
& {38.31}
& {0.9620}
& {34.07}
& {0.9213}
& {32.42}
& {0.9028}
& {33.10}
& {0.9370}
& {39.32}
& {0.9792}
\\
HAN~\cite{HAN} & $\times$2 & 63.6M & 14.6T%
& {38.27}
& {0.9614}
& {34.16}
& {0.9217}
& {32.41}
& {0.9027}
& {33.35}
& {0.9385}
& {39.46}
& {0.9785}  
\\
IGNN~\cite{IGNN} & $\times$2 & 49.5M & - %
& {38.24}
& {0.9613}
& {34.07}
& {0.9217}
& {32.41}
& {0.9025}
& {33.23}
& {0.9383}
& {39.35}
& {0.9786}
\\ 
NLSA~\cite{NLSA} & $\times$2 & 41.8M & 9.6T %
& 38.34 
& 0.9618 
& 34.08 
& {0.9231}
& 32.43 
& 0.9027 
& {33.42}
& {0.9394}
& {39.59}
& 0.9789
\\
DFSA+~\cite{DFSA} & $\times$2 & - & -%
& 38.38 
& 0.9620 
& 34.33 
& {0.9232}
& 32.50 
& 0.9036 
& {33.66}
& {0.9412}
& {39.98}
& 0.9798
\\
\rowcolor{tableblue}
\textbf{\algname{}}  & $\times$2  & 8.7M& 1.7T
& \blue{38.42}
& \blue{0.9622}
& \blue{34.40}
& \blue{0.9242}
& \blue{32.53}
& \blue{0.9043}
& \blue{33.73}
& \blue{0.9422}
& \blue{40.02}
& \blue{0.9801}
\\
\rowcolor{tableblue}
\textbf{\algname{}+}  \qquad\qquad\qquad\quad & $\times$2  & 8.7M &-
& \red{38.44}
& \red{0.9623}
& \red{34.49}
& \red{0.9248}
& \red{32.55}
& \red{0.9045}
& \red{33.86}
& \red{0.9430}
& \red{40.13}
& \red{0.9804}
\\
SwinIR$^\dagger$~\cite{SwinIR} & $\times$2  & 11.8M &2.3T
& {38.42}
& {0.9623}
& {34.46}
& {0.9250}
& {32.53}
& {0.9041}
& {33.81}
& {0.9427}
& {39.92}
& {0.9797}
\\
\whline
RCAN~\cite{RCAN} & $\times$3   & 15.6M& 1.6T
& {34.74}
&{0.9299}
& {30.65}
& {0.8482}
& {29.32}
& {0.8111}
& {29.09}
& {0.8702}
& {34.44}
&{0.9499}
\\
SAN~\cite{SAN} & $\times$3   & 15.9M& 1.6T
& {34.75}
& {0.9300}
& {30.59}
& {0.8476}
& {29.33}
& {0.8112}
& {28.93}
& {0.8671}
& {34.30}
& {0.9494}
\\
HAN~\cite{HAN}& $\times$3   & 64.3M & 6.5T
& {34.75}
& {0.9299}
& {30.67}
& {0.8483}
& {29.32}
& {0.8110}
& {29.10}
& {0.8705}
& {34.48}
& {0.9500}
\\
IGNN~\cite{IGNN} & $\times$3  & 49.5M&-
& {34.72}
& {0.9298}
& {30.66}
& {0.8484}
& {29.31}
& {0.8105}
& {29.03}
& {0.8696}
& {34.39}
& {0.9496}
\\
NLSA~\cite{NLSA}& $\times$3  & 44.7M& 4.6T
& 34.85 
& 0.9306 
& 30.70 
& 0.8485 
& 29.34 
& 0.8117 
& {29.25}
& {0.8726}
& 34.57 
& 0.9508
\\
DFSA+~\cite{DFSA} & $\times$3  & -& -
& \blue{34.92} 
& {0.9312}
& 30.83 
& 0.8507 
& 29.42
& 0.8128 
& {29.44}
& {0.8761}
& \blue{35.07} 
& 0.9525
\\
\rowcolor{tableblue} \textbf{\algname{}}   & $\times$3  & 8.7M &0.8T
& {34.91}
& \blue{0.9312}
& \blue{30.88}
& \blue{0.8514}
& \blue{29.43}
& \blue{0.8138}
& \blue{29.52}
& \blue{0.8782}
& {35.06}
& \blue{0.9526}
\\
\rowcolor{tableblue} \textbf{\algname{}+}   & $\times$3  &8.7M& -
& \red{34.97}
& \red{0.9315}
& \red{30.91}
& \red{0.8522}
& \red{29.47}
& \red{0.8144}
& \red{29.65}
& \red{0.8799}
& \red{35.21}
& \red{0.9533}
\\
SwinIR$^\dagger$~\cite{SwinIR} & $\times$3  & 11.9M& 1.0T
& {34.97}
& {0.9318}
& {30.93}
& {0.8534}
& {29.46}
& {0.8145}
& {29.75}
& {0.8826}
& {35.12}
& {0.9537}
\\
\whline
RCAN~\cite{RCAN} &$\times$4  & 15.6M& 0.9T
& {32.63}
& {0.9002}
& {28.87}
&{0.7889}
& {27.77}
& {0.7436}
&{26.82}
& {0.8087}
&{31.22}
& {0.9173}
\\ 
SAN~\cite{SAN} & $\times$4  & 15.9M& 0.9T
& {32.64}
& {0.9003}
& {28.92}
& {0.7888}
& {27.78}
& {0.7436}
& {26.79}
& {0.8068}
& {31.18}
& {0.9169}
\\
HAN~\cite{HAN} & $\times$4  & 64.2M&3.8T
& {32.64}
& {0.9002}
& {28.90}
& {0.7890}
& {27.80}
& {0.7442}
& {26.85}
& {0.8094}
& {31.42}
& {0.9177}
\\
IGNN~\cite{IGNN}  & $\times$4  & 49.5M&-
& {32.57}
& {0.8998}
& {28.85}
& {0.7891}
& {27.77}
& {0.7434}
& {26.84}
& {0.8090}
& {31.28}
& {0.9182}
\\
NLSA~\cite{NLSA} & $\times$4 & 44.2M&3.0T
& 32.59 
& 0.9000 
& 28.87 
& 0.7891 
& 27.78 
& 0.7444 
& {26.96}
& {0.8109}
& 31.27 
& 0.9184
\\
DFSA+~\cite{DFSA}  & $\times$4  & -&-
& {32.79}
& {0.9019}
& {29.06}
& {0.7922}
& {27.87}
& {0.7458}
& {27.17}
& {0.8163}
& {31.88}
& \red{0.9266}
\\
\rowcolor{tableblue}
\textbf{\algname{}}   & $\times$4  & 8.7M&0.4T
& \blue{32.81}
& \blue{0.9024}
& \blue{29.07}
& \blue{0.7934}
& \blue{27.90}
& \blue{0.7477}
& \blue{27.26}
& \blue{0.8197}
& \blue{31.92}
& {0.9230}
\\
\rowcolor{tableblue}
\textbf{\algname{}+}   & $\times$4  & 8.7M&-
& \red{32.87}
& \red{0.9030}
& \red{29.12}
& \red{0.7941}
& \red{27.93}
& \red{0.7483}
& \red{27.39}
& \red{0.8223}
& \red{32.13}
& \blue{0.9248}
\\
SwinIR$^\dagger$~\cite{SwinIR}  & $\times$4  & 11.9M & 0.6T
& {32.92}
& {0.9044}
& {29.09}
& {0.7950}
& {27.92}
& {0.7489}
& {27.45}
& {0.8254}
& {32.03}
& {0.9260}
\\
\whline             
\end{tabular}
\end{center}
\end{table*}

\begin{figure*}[!t]
  \tabcolsep=1.0pt
  \centering
  \small
  \fontsize{8.5pt}{9.5pt}\selectfont
  \begin{tabular}{ccccccc}
  \multirow{-4.2}{*}{\includegraphics[width=.21\linewidth, height=.175\linewidth]{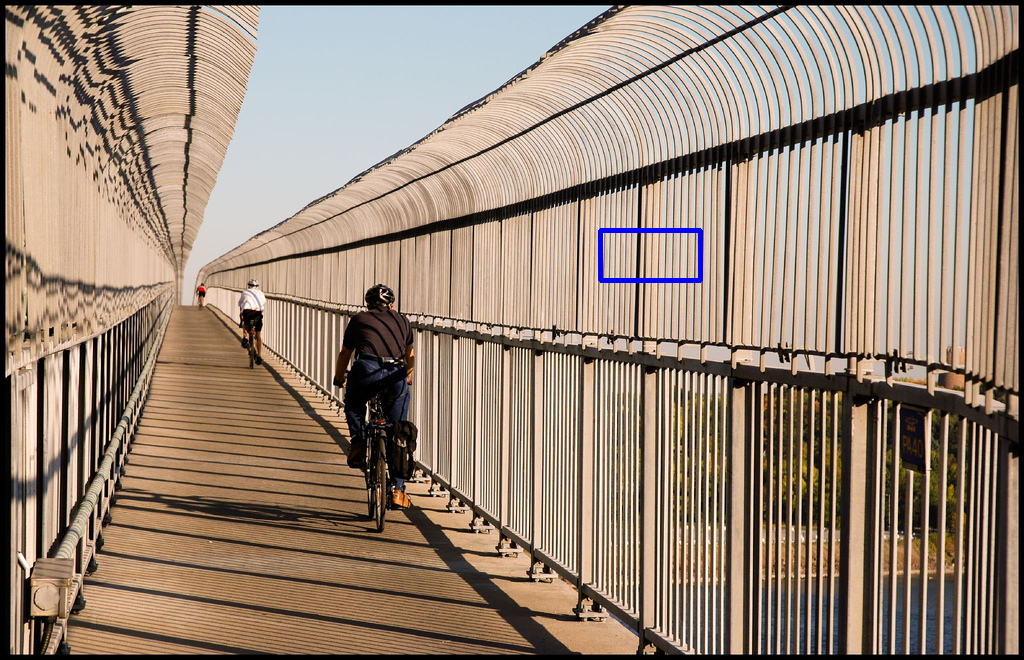}} & 
  \includegraphics[width=.15\linewidth, height=.075\linewidth]{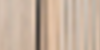}& 
  \includegraphics[width=.15\linewidth, height=.075\linewidth]{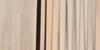}& \includegraphics[width=.15\linewidth, height=.075\linewidth]{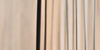}& \includegraphics[width=.15\linewidth, height=.075\linewidth]{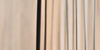}& 
  \includegraphics[width=.15\linewidth, height=.075\linewidth]{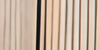} \\
   & Bicubic & EDSR~\cite{EDSR} & RCAN~\cite{RCAN} & SAN~\cite{SAN} & HAN~\cite{HAN}\\
    & \includegraphics[width=.15\linewidth, height=.075\linewidth]{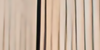}& \includegraphics[width=.15\linewidth, height=.075\linewidth]{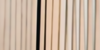}& \includegraphics[width=.15\linewidth, height=.075\linewidth]{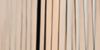}& 
    \includegraphics[width=.15\linewidth, height=.075\linewidth]{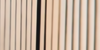}
    & \includegraphics[width=.15\linewidth, height=.075\linewidth]{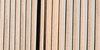}\\
    \emph{img\_024} from Urban100~\cite{Urban100} & IGNN~\cite{IGNN} & DFSA~\cite{DFSA} & SwinIR~\cite{SwinIR} & \red{MAN} & HR\\
  \multirow{-4.2}{*}{\includegraphics[width=.21\linewidth, height=.175\linewidth]{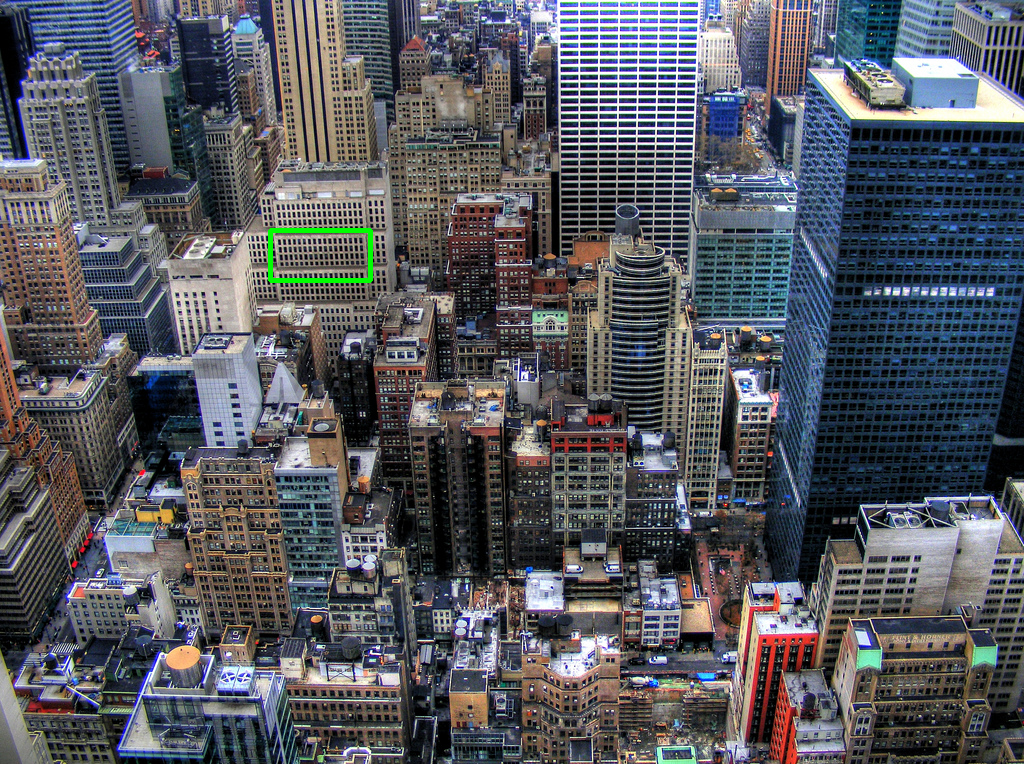}} & 
  \includegraphics[width=.15\linewidth, height=.075\linewidth]{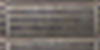}& 
  \includegraphics[width=.15\linewidth, height=.075\linewidth]{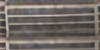}& \includegraphics[width=.15\linewidth, height=.075\linewidth]{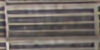}& \includegraphics[width=.15\linewidth, height=.075\linewidth]{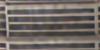}& 
  \includegraphics[width=.15\linewidth, height=.075\linewidth]{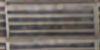} \\
   & Bicubic & EDSR~\cite{EDSR} & RCAN~\cite{RCAN} & SAN~\cite{SAN} & HAN~\cite{HAN}\\
   & \includegraphics[width=.15\linewidth, height=.075\linewidth]{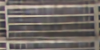}& \includegraphics[width=.15\linewidth, height=.075\linewidth]{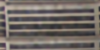}& \includegraphics[width=.15\linewidth, height=.075\linewidth]{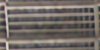}& 
    \includegraphics[width=.15\linewidth, height=.075\linewidth]{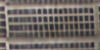}
    & \includegraphics[width=.15\linewidth, height=.075\linewidth]{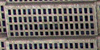}\\
    \emph{img\_073} from Urban100~\cite{Urban100} & IGNN~\cite{IGNN} & DFSA~\cite{DFSA} & SwinIR~\cite{SwinIR} & \red{MAN} & HR\\
\end{tabular}
  \caption{Visual comparison for classical SR models with an upscaling factor $\times$4.}
  \label{tab:Bi_fig2}
  \vspace{-1mm}
  \end{figure*}

\subsection{Comparisons with classical SR models}
To validate the effectiveness of our MAN, we compare our normal model to several SOTA classical ConvNets~\cite{RCAN,SAN,IGNN,HAN,NLSA,DFSA}. We also add SwinIR~\cite{SwinIR} for reference. In \cref{tab:sr_results}, the quantitative results show that our MAN exceeds other convolutional methods to a large extent. The maximum improvement on PSNR reaches 0.69\,dB for $\times$2, 0.77\,dB for $\times$3, and 0.81\,dB for $\times$4. Moreover, we compare our MAN with SwinIR. For $\times$2, our MAN achieves competitive or even better performance than SwinIR. The PSNR value on Manga109 is boosted from 39.92\,dB to 40.02\,dB. For $\times$4, MAN is slightly behind SwinIR because the latter uses the $\times$2 model as the pre-trained model. More importantly, MAN is significantly smaller than existing methods. 

In \cref{tab:Bi_fig2}, we also visualize the qualitative results of several models on the Urban100 ($\times$4) benchmark dataset. For \emph{img\_024}, compared with other models generating the distorted fence, our MAN rebuilds a clear structure from the blurred input. Similarly, in \emph{img\_073}, MAN is the only model that restores the windows of the building.


\begin{table*}[!t]
\center
\begin{center}
\caption{Quantitative comparison (average PSNR/SSIM) with state-of-the-art approaches for \textbf{\underline{tiny/light image SR}} on benchmark datasets ($\times$4). The best and second best performances are \red{highlighted} and \blue{underlined}, respectively.}
\label{tab:tiny/light_sr_results} 
\small
\fontsize{8.5pt}{9.5pt}\selectfont
\tabcolsep=3.1pt
\begin{tabular}{l|c|c|c|c|c|c|c|c|c|c|c|c|c}
\whline
\multirow{2}{*}{Method} & \multirow{2}{*}{Scale} & \multirow{2}{*}{\#Params} & \multirow{2}{*}{\#FLOPs} &  \multicolumn{2}{c|}{Set5~\cite{Set5}} &  \multicolumn{2}{c|}{Set14~\cite{Set14}} &  \multicolumn{2}{c|}{BSD100~\cite{B100}} &  \multicolumn{2}{c|}{Urban100~\cite{Urban100}} &  \multicolumn{2}{c}{Manga109~\cite{Manga109}}  
\\
\cline{5-14}
&  &  &  & PSNR & SSIM & PSNR & SSIM & PSNR & SSIM & PSNR & SSIM & PSNR & SSIM 
\\
\whline
FSRCNN~\cite{FSRCNN} & $\times$4  & 12K & 4.6G
& 30.71 
& 0.8657
& 27.59
& 0.7535
& 26.98
& 0.7150
& 24.62
& 0.7280
& 27.90
& 0.8517
\\
LAPAR-C~\cite{LAPAR} & $\times$4  & 115K & 25.0G
& {31.72}
& 0.8884
& 28.31
& 0.7740
& {27.40}
& 0.7292
& {25.49}
& 0.7651
& {29.50}
& {0.8951}
\\
ECBSR-M10C32~\cite{ECB} & $\times$4  & 98K & 5.7G
& 31.66
& {0.8911}
& {28.15}
& {0.7776}
& {27.34}
& \red{0.7363}
& {25.41}
& \blue{0.7653}
& -
& -
\\
ShuffleMixer-tiny~\cite{shufflemixer} & $\times$4  & 113K & 8.0G
& \blue{31.88}
& \blue{0.8912}
& \blue{28.46}
& \blue{0.7779}
& \blue{27.45}
& {0.7313}
& \blue{25.66}
& \blue{0.7690}
& \blue{29.96}
& \blue{0.9006}
\\
ETDS-L~\cite{ETDS} & $\times$4  & 170K & 9.8G
& {31.69}
& {0.8889}
& {28.31}
& {0.7751}
& {27.37}
& {0.7302}
& {25.47}
& {0.7643}
& -
& -
\\
\rowcolor{tableblue}
\textbf{\algname{}-tiny}   & $\times$4  &  150K & {8.4G} %
& \red{32.07}
& \red{0.8930}
& \red{28.53}
& \red{0.7801}
& \red{27.51}
& \blue{0.7345}
& \red{25.84}
& \red{0.7786}
& \red{30.18}
& \red{0.9047}
\\
EDSR-baseline~\cite{EDSR} & $\times$4  & 1518K & {114G} %
& {32.09}
& {0.8938}
& {28.58}
& {0.7813}
& {27.57}
& {0.7357}
& {26.04}
& {0.7849}
& {30.35}
& {0.9067}
\\
\whline
IMDN~\cite{IMDN} & $\times$4  & 715K & 40.9G
& 32.21
& 0.8948
& 28.58
& 0.7811
& 27.56
& 0.7353 
& 26.04
& 0.7838
& 30.45
& 0.9075
\\
LatticeNet~\cite{LatticeNet} & $\times$4  & 777K & 43.6G
& {32.30}
& {0.8962}
& {28.68}
& {0.7830}
& {27.62}
& {0.7367}
& {26.25}
& {0.7873}
& -
& -
\\
DIPNet~\cite{DIPNet} & $\times$4  & 543K & -
& {32.20}
& {0.8950}
& {28.58}
& {0.7811}
& {27.59}
& {0.7364}
& {26.16}
& {0.7879}
& 30.53
& 0.9087
\\
SwinIR-light~\cite{SwinIR} & $\times$4  & 897K & {49.6G} %
& \blue{32.44}
& \blue{0.8976}
& {28.77}
& \blue{0.7858}
& \blue{27.69}
& \blue{0.7406}
& {26.47}
& {0.7980}
& \blue{30.92}
& \blue{0.9151}
\\
ELAN-light~\cite{ELAN} & $\times$4  & 601K & 43.2G
& {32.43}
& {0.8975}
& \blue{28.78}
& \blue{0.7858}
& \blue{27.69}
& \blue{0.7406}
& \blue{26.54}
& \blue{0.7982}
& \blue{30.92}
& {0.9150}
\\
\rowcolor{tableblue}
\textbf{\algname{}-light}  \qquad\qquad\qquad\quad  & $\times$4  & 840K & {47.1G} %
& \red{32.50}
& \red{0.8988}
& \red{28.87}
& \red{0.7885}
& \red{27.77}
& \red{0.7429}
& \red{26.70}
& \red{0.8052}
& \red{31.25}
& \red{0.9170}
\\
EDSR~\cite{EDSR} & $\times$4  & 43090K & {2895G} %
& {32.46}
& {0.8968}
& {28.80}
& {0.7876}
& {27.71}
& {0.7420}
& {26.64}
& {0.8033 }
& {31.02}
& {0.9148}\\
\whline  
\end{tabular}
\end{center}
\end{table*}

\begin{figure*}[!t]
\small
\scriptsize
\tabcolsep=1pt
  \centering
  \begin{tabular}{ccccccccccc} 
  {\rotatebox{90}{\emph{img\_078}}}
  &\includegraphics[width=.094\linewidth, height=.11\linewidth]{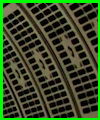}
  &\includegraphics[width=.094\linewidth, height=.11\linewidth]{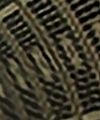} 
  &\includegraphics[width=.094\linewidth, height=.11\linewidth]{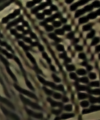} 
  &\includegraphics[width=.094\linewidth, height=.11\linewidth]{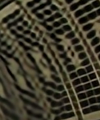}
  &\includegraphics[width=.094\linewidth, height=.11\linewidth]{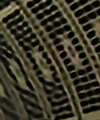}
  &\includegraphics[width=.094\linewidth, height=.11\linewidth]{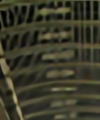}
  &\includegraphics[width=.094\linewidth, height=.11\linewidth]{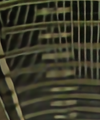}
  &\includegraphics[width=.094\linewidth, height=.11\linewidth]{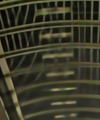} 
  &\includegraphics[width=.094\linewidth, height=.11\linewidth]{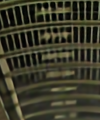} 
  &\includegraphics[width=.094\linewidth, height=.11\linewidth]{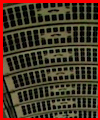}\\
  {\rotatebox{90}{\emph{img\_092}}}
  &\includegraphics[width=.094\linewidth, height=.11\linewidth]{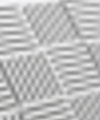} 
  &\includegraphics[width=.094\linewidth, height=.11\linewidth]{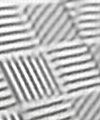}
  &\includegraphics[width=.094\linewidth, height=.11\linewidth]{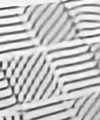} 
  &\includegraphics[width=.094\linewidth, height=.11\linewidth]{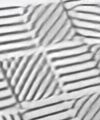} 
  &\includegraphics[width=.094\linewidth, height=.11\linewidth]{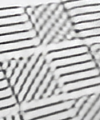} 
  &\includegraphics[width=.094\linewidth, height=.11\linewidth]{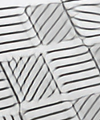}
  &\includegraphics[width=.094\linewidth, height=.11\linewidth]{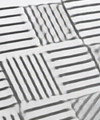}
  &\includegraphics[width=.094\linewidth, height=.11\linewidth]{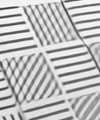} 
  &\includegraphics[width=.094\linewidth, height=.11\linewidth]{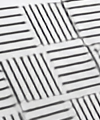} 
  &\includegraphics[width=.094\linewidth, height=.11\linewidth]{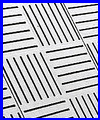}\\
  & HR/Bicubic & FSRCNN~\cite{FSRCNN} & LAPAR-C~\cite{LAPAR} & VDSR~\cite{VDSR} & \textbf{MAN-tiny} & IMDN~\cite{IMDN} & SwinIR-L~\cite{SwinIR} & EDSR~\cite{EDSR} & \textbf{MAN-light}& HR \\
\end{tabular}
  \caption{Visual comparison for tiny/lightweight SR models with an upscaling factor $\times$4.}
  \label{tab:Bi_fig}
  \vspace{-2mm}
  \end{figure*}

\subsection{Comparisons with tiny/light SR models}
To verify the efficiency and scalability of our MAN, we compare MAN-tiny and MAN-light to some state-of-the-art tiny~\cite{FSRCNN,LAPAR,ECB,shufflemixer,EDT} and lightweight~\cite{IMDN,LatticeNet,DIPNet,SwinIR,ELAN} SR models. \cref{tab:tiny/light_sr_results} presents the numerical results that our MAN-tiny/light outperforms all other tiny/lightweight methods. Specifically, MAN-tiny exceeds second place by about 0.2\,dB on Set5, Urban100, and Manga109, and around 0.07\,dB on Set14 and BSD100. We also list EDSR-baseline~\cite{EDSR} for reference. Our tiny model has less than 150K parameters but achieves a similar restoration quality with EDSR-baseline, which is 10$\times$ larger than ours. Similarly, our MAN-light surpasses both CNN-based and transformer-based SR models. In comparison with IMDN (CNN) and SwinIR-light/ELAN-light (Transformer), our model leads by 0.66\,dB/0.23\,dB on Urban100 ($\times$4) benchmark. Moreover, our MAN-light is superior to traditional performance-oriented EDSR. In detail, the proposed model takes only 2\% of the parameters and computations of EDSR while having high PSNR on all benchmarks. 

In \cref{tab:Bi_fig}, we also exhibit the visual results of several tiny/lightweight models on Urban100 ($\times$4). For \emph{img\_078}, the tiny and light models are tested with the patches framed by {\color{green}green} and {\color{red}red} boxes, respectively. Generally, MANs can restore the texture better and clearer than other methods.

\begin{table}[t]
\center
\begin{center}
\small
\fontsize{8.5pt}{9.5pt}\selectfont
\tabcolsep=3pt
\begin{tabular}{l|c|c|c|c|c|c}
\whline
{Method} &
{\#Params} & {\#FLOPs} &  {Set5} & {B100}& {U100} &  {M109}   \\
\whline
IPT~\cite{IPT} & 115.5M &- 
& 38.37 
& 32.48
& 33.76
& - 
\\
EDT-B~\cite{EDT} & 11.5M & {37.6G} %
& {38.45}
& {32.52}
& {33.80}
& {39.93}
\\
HAT~\cite{HAT} & 20.8M & {103.7G}$^\dagger$ %
& {38.63}
& {32.62}
& {34.45}
& {40.26}
\\
DAT~\cite{DAT} & 14.7M & {245.4G}$^\dagger$ %
& {38.58}
& {32.61}
& {34.37}
& {40.33}
\\
\rowcolor{tableblue}
MAN \qquad\qquad\quad & 8.7M & {19.8G} %
& {38.42}
& {32.53}
& {33.73}
& {40.02}
\\
\whline
\end{tabular}
\end{center}
\caption{Quantitative comparison with sota transformers  ($\times$2). \#FLOPs are calculated with 48$\times$48/64$\times$64$^\dagger$ inputs.}
\label{tab:sota_tramsformer}
\end{table}

\subsection{Comparisons with SR Transformers}
Although MANs achieve remarkable improvement compared to existing ConvNet-based models, more comparison with some transformer-based approaches is essential. In \cref{tab:sota_tramsformer}, we include the competitive IPT~\cite{IPT}, EDT~\cite{EDT}, HAT~\cite{HAT}, and DAT~\cite{DAT} for discussion. MAN achieves similar quality as EDT-B with only 75\% params and 52\% FLOPs (48$\times$48 input). The HAT and DAT are much larger models than EDT or MAN, which perform superior to both. In a word, MAN can perform on par with or even better than these transformer-based methods (SwinIR, EDT) with similar model sizes, showing ConvNet's vitality in low-level.

\section{Conclusion}
This paper proposes a multi-scale attention network (MAN) for super-resolution under multiple complexities. MAN adopts transformer-style blocks for better modeling representation. To effectively and flexibly establish long-range correlations among various regions, we develop multi-scale large kernel attention (MLKA) that combines large kernel decomposition and multi-scale mechanisms. Furthermore, we propose a simplified feed-forward network (GSAU) that integrates gate mechanisms and spatial attention to activate local information and reduce model complexity. Extensive experiments have demonstrated that our CNN-based MAN can achieve better performance than previous SOTA ConvNets and keep pace with transformer-based methods in a more efficient manner.
{
    \small
    \bibliographystyle{ieeenat_fullname}
    \bibliography{main}
}


\end{document}